\documentclass[%
reprint,
aps,
prb,
amsmath,
twocolumn,
amssymb,
floatfixng,
showpacs,
superscriptaddress,
footinbib,
multicol
]{revtex4-2}
\usepackage{graphicx}
\usepackage{epstopdf}
\usepackage{dcolumn}
\usepackage{bm}
\usepackage{color}
\usepackage{xcolor}
\usepackage{amsmath}
\usepackage{braket}
\usepackage{amsfonts}
\usepackage{soul}
\usepackage{url}
\usepackage{array,tabularx}
\usepackage{graphics}
\usepackage{times}
\usepackage{subfigure}
\usepackage[export]{adjustbox}
\usepackage[colorlinks=true,linktoc=page,linkcolor=red,citecolor=blue]{hyperref}

\definecolor{OliveGreen}{rgb}{0,0.6,0}

\begin{document}

\title{
Crossovers from nonlinear wave-packet acceleration to wave-mixing and self-trapping in the Hatano–Nelson model
}

\author{Bertin Many Manda}
\email{bmany@tauex.tau.ac.il}
\affiliation{School of Mechanical Engineering, Tel Aviv University, Tel Aviv 69978, Israel}
\affiliation{Laboratoire d’Acoustique de l’Universit\'e du Mans (LAUM),
UMR 6613, Institut d’Acoustique - Graduate School (IA-GS), CNRS,
Le Mans Universit\'e, Av. Olivier Messiaen, 72085 Le Mans, France}
\author{Vassos Achilleos}
\email{vassos.achilleos@univ-lemans.fr}
\affiliation{Laboratoire d’Acoustique de l’Universit\'e du Mans (LAUM),
UMR 6613, Institut d’Acoustique - Graduate School (IA-GS), CNRS,
Le Mans Universit\'e, Av. Olivier Messiaen, 72085 Le Mans, France}

\begin{abstract}
We demonstrate that wave amplification enables even weak nonlinearities to reshape linear wave-packet transport in nonreciprocal systems. 
We study the dynamics of bulk Gaussian wave-packets in the Hatano--Nelson model with on-site cubic nonlinearity. 
We show that the interplay between nonlinearity and amplification generates growing frequency shifts that drive the wave-packet through three successive dynamical regimes: an early nonlinear-skin regime with coherent propagation, an intermediate wave-mixing regime driven by eigenmode resonances, and a self-trapping regime in which part of the packet localizes while the remainder ballistically spreads along the system favored direction. 
The crossover time scales are set by the width and averaged spacing of the eigenfrequency spectrum. 
Crucially, within the nonlinear-skin regime, we derive analytical predictions for the wave-packet dynamics and show that nonlinearity couples amplification, dispersion, and nonreciprocity, thereby modifying the magnitude of the wave-packet acceleration and introducing an explicit time dependence into its evolution.
Focusing nonlinearities suppress the acceleration and cause it to decrease in time, whereas defocusing nonlinearities enhance it and cause it to increase. We further show that nonlinear interactions typically break down the wave-packet before the non-Hermitian jump can occur. Our results provide a route toward accurate control of waves in nonreciprocal metamaterials.
\end{abstract}

\maketitle

\maketitle


\section{Introduction}
In just over a decade, nonreciprocal systems have emerged as an interdisciplinary research field spanning a wide range of physical platforms~\cite{GBVC2020,HHIAKMLSGT2020,WKHHSGTS2020,LSMZYWJCZ2021,ZYGGCYCXLJYSCZ2021,WWM2022,ZLYC2022,WWLQZLLL2025,LM2026,RB2026}. 
Regardless of the specific implementation, two paradigmatic models--the Hatano-Nelson (HN) model~\cite{HN1996,HN1998,ASPMCACBOPG2023} and Klein-Gordon (KG) chain~\cite{GBVC2020,M2026} of asymmetrically coupled classical oscillators--have emerged as providing a common framework for understanding wave dynamics in such systems. 
A hallmark property of these models is the non-Hermitian skin effect (NHSE), characterized by an exponential localization of eigenmodes at a single end of the system under open boundary conditions (OBCs)~\cite{KEBB2018,LT2019,YW2018,BKS2020,OKSS2020,LTLL2023,OS2023,WC2023,ZZCLC2023}. 
The importance of the NHSE for topological wave transport has been widely discussed~\cite{LBHCN2017,GAKTHU2018,SZF2018,LAZV2019,AGU2020,BKS2020,OKSS2020,OS2023}, and supported by numerous experimental observations in mechanical~\cite{GBVC2020}, photonic~\cite{WWLQZLLL2025}, atomic~\cite{LXDLLGYY2022}, electrical~\cite{PJYW2022}, and acoustic~\cite{MAPPA2023} platforms.

\begin{figure}[!tb]
    \centering
    \includegraphics[width=\columnwidth]{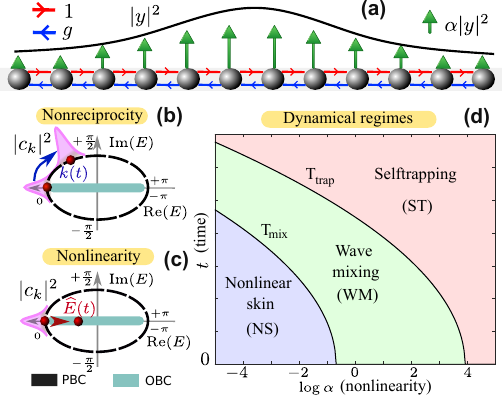}
    \caption{
        (a) Schematic of the nonlinear HN chain. 
        The parameters $g$ and $\alpha$ control the strengths of the nonreciprocity and nonlinearity, respectively.
        (b) Effect of nonreciprocity on an initial wave-packet centered at $k_0=0$ in the eigenmode space. 
        (c) Same as (b), but illustrating the effect of nonlinearity with $\widehat{E}$ being the renormalized nonlinear frequency.
        (d) Theoretical dynamical regimes of nonlinear wave-packets in the HN model: (blue) nonlinear-skin, (green) wave-mixing, and (red) self-trapping regimes. 
        The parameter space shows the nonlinear coefficient $\alpha$ against time $t$ with the boundaries, $T_{\mathrm{mix}}$ and $T_{\mathrm{trap}}$ obtained by comparing the nonlinear frequency shift $\nu(t)$ to the linear eigenfrequency scales, $d$ and $\Delta$ (see text for details).
    }
    \label{fig:schematic}
\end{figure}

Recent research in these systems has begun looking at their dynamical properties~\cite{E2022,L2022,JCZL2023,KMM2023,MCKA2024,BWHLC2025,JMAFS2025,L2025,WWLQZLLL2025}. 
In particular, in the linearized limit of the lattices above, dynamical signatures of the NHSE, like unidirectional transport and the dynamical skin effect, have been demonstrated both theoretically~\cite{LW2022,BN2024,Y2026} and experimentally~\cite{LXDLLGYY2022,LWWLMJ2024}. 
Regarding the propagation of localized wave-packets, their asymptotics have also attracted significant attention. 
In this context, it was shown that their rates of growth depend on the interplay between dispersion and nonreciprocity. 
Moreover, this precise interplay also gives rise to another remarkable phenomenon: the packet center of mass exhibits a time-dependent acceleration, in contrast to uniform motion of Hermitian pulses~\cite{L2022,JMAFS2026,HO2025}. 
Even more striking, this center of mass can also abruptly relocate to a different position, a phenomenon dubbed as non-Hermitian wave jump~\cite{HO2025,JMAFS2026}. 
While wave acceleration has recently been demonstrated experimentally in nonreciprocal photonic systems~\cite{XLWXLY2024}, the observation of the non-Hermitian wave jump remains an open challenge.

On the other hand, the exponential growth of waves in nonreciprocal systems can rapidly amplify even weak nonlinear effects, potentially leading to pulse dynamics that differ drastically from the predictions of the linear theory~\cite{WWM2022,VGGSMC2024,VGBVTCC2025,WWLQZLLL2025}. 
Nevertheless, most existing studies, including those cited above, remain restricted to the linear limit. 
Indeed, only a few works have addressed localized wave-packets in nonlinear nonreciprocal models~\cite{E2022,WWLQZLLL2025}. 
In particular, Ref.~\cite{E2022} considers single-site excitations, relying on numerical simulations of small lattices  and over short times. 
Consequently, it does not provide a sufficient understanding of the fundamental mechanisms governing wave-packet dynamics in nonlinear nonreciprocal systems. 
This leaves several important open questions: How does nonlinearity modify wave-packet transport in nonreciprocal systems? 
Under what conditions does it suppress coherent propagation and instead induce wave-mixing or self-trapping? 
And how does it reshape the wave acceleration and non-Hermitian jump?

In this paper, we carefully analyze the time evolution of initially localized wave-packets launched at the center of the HN lattice in the presence of on-site cubic nonlinearity, Fig.~\ref{fig:schematic}(a). 
At first glance, this problem involves understanding the interplay between nonreciprocity, dispersion, and nonlinearity. 
Considering these mechanisms separately, nonreciprocity causes the eigenmode center of mass to drift due to a gradient in their amplification~\cite{HO2025,JMAFS2026}, as illustrated by the blue arrow in Fig.~\ref{fig:schematic}(b). 
On the other hand, finite amplitude waves produce a nonlinear frequency shift which detunes eigenmodes away from the linear spectrum, as indicated by the red arrow in Fig.~\ref{fig:schematic}(c). 
One of the main findings of this work is that nonlinearity not only introduces an explicit time dependence into the wave-packet acceleration but also typically destroys the pulse coherence before a non-Hermitian jump can occur.

The paper is organized as follows. 
In Sec.~\ref{sec:model}, we introduce the nonlinear HN chain. 
In Sec.~\ref{sec:secular_form}, we derive the secular-form equations and identify the expected dynamical regimes. 
Section~\ref{sec:analytic_spreading} develops analytical predictions for the evolution of Gaussian wave-packets, while Sec.~\ref{sec:regime_transition} provides estimates of the corresponding characteristic crossover times.
In Sec.~\ref{sec:numerical_results}, we present numerical results for the nonlinear HN chain. 
Finally, Sec.~\ref{sec:conclusion} summarizes our findings, presents our conclusions, and outlines several directions for future research.
Further details of the analytical derivations and additional results from numerical simulations are collected in the appendices.

\section{\label{sec:model}Model}
Our system is described by the nonlinear HN chain with on-site self-interaction potentials~\cite{MCKA2024,WWLQZLLL2025},
\begin{equation}
    -i\frac{dy_n}{dt} = gy_{n-1} + y_{n+1} + \alpha |y_n|^2 y_n.
    \label{eq:equation_motion_discrete}
\end{equation}
Here the $dy_n/dt=\dot{y}_n$ denotes the time derivative of the complex amplitude $y_n$ of oscillator with index $n$, $g$ and $\alpha$ represent the strengths of nonreciprocity and nonlinearity, and $i^2=-1$. 
This system is reciprocal at $g=1$ and nonreciprocal for $g>1$.
On the other hand, it is linear for $\alpha=0$ and nonlinear otherwise with focusing and defocusing nonlinearities corresponding to $\alpha>0$ and $\alpha<0$ values respectively~\cite{K2009}. 
Experimental realizations of such systems have already been reported, for example, in optical waveguides~\cite{WWLQZLLL2025}.
Throughout this work, we consider large lattices.

We implement bulk Gaussian wave-packets at time $t=0$, of width $\sigma_0$, centered at site $n_0=0$, and with its wavenumber distribution clustered around $k_0$~\cite{CDL1986,S2016,GS2018},
\begin{equation}
    y_n(0)=\frac{e^{ik_0 n}}{\sqrt[4]{2\pi\sigma_0^2}}\exp\left[-\frac{(n-n_0)^2}{4\sigma_0^2} \right].
    \label{eq:initial_condition_discrete}
\end{equation}
Without loss of generality we set $k_0=0$.

In the linear limit ($\alpha=0$), the HN chain of size $N$ with periodic boundary conditions (PBCs) admits solutions of the form $y_n(t)=u_{n}e^{iEt}$~\cite{MCKA2024}. 
Substituting this ansatz into Eq.~\eqref{eq:equation_motion_discrete} yields a linear eigenvalue problem, where $E$ has units of frequency and $u_n$ denotes the corresponding wavefunction.
We find the eigenfrequency spectrum,
\begin{equation}
    E_k = \varepsilon_k - i\Omega_k, \quad 
    \label{eq:specrum}
\end{equation}
with
\begin{equation}
 \varepsilon_k = (g+1)\cos k,~ \Omega_k = (g-1)\sin k,~ k = \frac{2\pi j}{N}.
\end{equation}
and $j = 1,2,\ldots,N$.
We refer to $k$ as wavenumber.
It is these complex eigenfrequencies which are at the origin of wave amplification and attenuation in the HN chain~\cite{HO2025,JMAFS2026}.
It follows that the width of the eigenfrequency spectrum along the real axis in the complex plane [Figs.~\ref{fig:schematic}(b)--(c)] reads
\begin{equation}
    \Delta = \lvert \varepsilon_{k=0}-\varepsilon_{k=\pi}\rvert = 2(g+1).
    \label{eq:width_frequency}
\end{equation}

The eigenmodes, $u_{k,n}=e^{ikn}/\sqrt{N}$ associated with $E_k$ are extended for all $k$~\cite{KIKIPO}. 
Indeed, for ordinary exponential localization, the eigenmode density profile satisfies $\lvert u_{k,n}\rvert^2 = e^{-\lvert n\rvert/\zeta_k}$,
where $\zeta_k$ is referred to as the localization length~\cite{KM1993,IKM2012}.
This length characterizes the spatial extent of the eigenmodes within the lattice.
Consequently, the participation number $P_k=\left(\sum_n \lvert u_{k,n}\rvert^2\right)^2/\sum_n \lvert u_{k,n}\rvert^4$, which measures the number of significantly occupied sites in the norm density profile $|u_{k,n}|^2$, provides a good estimate of $\zeta_k$. 
We find that all eigenmodes have the same participation number, $P_k=N$, meaning that they all extend over the entire lattice.
It follows that the average frequency spacing along the real axis between eigenmodes within the same localization length leads to
\begin{equation}
    d = \frac{\Delta}{P_k}, \quad \mbox{ where} \quad  P_k = N.
    \label{eq:average_spacing}
\end{equation}
These two frequency scales, $d$ [Eq.~\eqref{eq:average_spacing}] and $\Delta$ [Eq.~\eqref{eq:width_frequency}] are expected to determine the detailed evolution of the wave-packet in the presence of nonlinearity.

It is also useful to write the equations of motion of the nonlinear HN model [Eq.~\eqref{eq:equation_motion_discrete}] in eigenmode space. 
To this end, we expand the amplitude $y_n$ in the eigenmode basis $u_{k,n}=e^{ikn}/\sqrt{N}$ assuming orthonormalization.
Using the transformations $y_n= \sum c_k e^{ikn}/\sqrt{N}$ and $c_k = \sum y_n e^{-ikn}/\sqrt{N}$, we obtain the equations of motion for the eigenmode variables $c_k$,
\begin{equation}
    -i\frac{dc_k}{dt} = E_k c_k + \frac{\alpha}{N} \sum_{k_1,k_2,k_3} c_{k_1}c_{k_2}^\star c_{k_3} \delta_{k_1-k_2+k_3-k,~0}
    \label{eq:eqs_mot_nonlinear_NMspace}
\end{equation}
where $E_k$ [Eq.~\eqref{eq:specrum}] is the frequency of the eigenmode with wavenumber $k$, $c_k $ determines its complex time-dependent amplitude and $\{{ }^\star\}$ denotes the complex conjugate operation.
Further, $\delta_{k_1-k_2+k_3-k,\,0}$ stands for the Kronecker delta, which is equal to one when the condition
\begin{equation}
    k_1-k_2+k_3-k = 0,
    \label{eq:resonance_conditions}
\end{equation}
is satisfied and zero otherwise, with the above relation understood modulo $2\pi$.
It follows that in the presence of nonlinearity the $\delta_{k_1-k_2+k_3-k,\,0}$ induces an overlap between the eigenmodes.

Bearing the above in mind, we see that the nonlinear frequency shift of a single oscillator is $\nu_n(t)=\alpha \lvert y_n(t)\rvert^2$ [see Eq.~\eqref{eq:equation_motion_discrete}] while the one of a single eigenmode reads $\nu_k(t)=\alpha \lvert c_k(t)\rvert^2/N$ [see Eq.~\eqref{eq:eqs_mot_nonlinear_NMspace}].

\section{\label{sec:secular_form}Secular form and expected dynamical regimes}

In this section, we derive the secular form of the nonlinear HN dynamics and use it to identify the expected dynamical regimes of wave-packet evolution. 
This analysis aims to clarify whether and how on-site nonlinearity induces interactions between the eigenmodes $u_{k,n}=e^{ikn}/\sqrt{N}$, in the presence of non-Hermitian amplification and attenuation. 
To the best of our knowledge, this mechanism has not been addressed in previous studies, including Ref.~\cite{PRJKC2022}.

\subsection{\label{subsec:secular_form}Secular form equations and nonlinear frequency shift}

Let us introduce the simple transformation $c_k(t)=\phi_k(t)e^{\Omega_k t}e^{i\varepsilon_k t}$
and substitute it into Eq.~\eqref{eq:eqs_mot_nonlinear_NMspace}.
This leads to,
\begin{equation}
\begin{split}
    -i\frac{d\phi_k}{dt}
    = \frac{\alpha}{N}
    &\sum_{k_1,k_2,k_3} 
    \phi_{k_1}\phi_{k_2}^\star\phi_{k_3}
    e^{(\Omega_{k_1}+\Omega_{k_2}+\Omega_{k_3}-\Omega_k ) t} \times \ldots \\
    & \times e^{i (\varepsilon_k+\varepsilon_{k_2}-\varepsilon_{k_1}-\varepsilon_{k_3}) t}
    \delta_{k_1-k_2+k_3-k,0}.
\end{split}
    \label{eq:before_secular_form}
\end{equation}
Clearly in the absence of nonlinearity, $\alpha = 0$, we find $\dot{\phi_k} = 0$, resulting in all eigenmode norm densities being constants of motion of the HN chain, i.e. $\lvert \phi_k (t)\rvert^2 = \lvert C_k\rvert^2$ with $\phi_k (0) = C_k$.
Physically, this means that eigenmodes do not interact in the linear limit of the HN chain.
They solely amplify or attenuate as time evolves depending on the values of $ \Omega_k$,  since 
\begin{equation}
    c_k (t) = C_k e^{\Omega_k t} e^{i\varepsilon_k t},
    \label{eq:solution_eigenmodes_space}
\end{equation}

In this context, it is useful to briefly review the basics of wave-packet phenomena identified in previous studies~\cite{HO2025,JMAFS2026}. 
In the eigenmode space, the initial pulse of Eq.~\eqref{eq:initial_condition_discrete} reads
\begin{equation}
    c_k(t=0) \equiv C_k, \quad C_k = C_0 e^{-\sigma_0^2 k^2},
    \label{eq:initial_conditions_mode}
\end{equation}
and its subsequent time evolution is obtained from Eq.~\eqref{eq:solution_eigenmodes_space}. 
Here $C_0$ is a complex constant with its subscript denoting the wavenumber $k=0$, at which the wave-packet is centered in the eigenmode basis.
Thus the $\Omega_k$ not only controls the overall amplification or attenuation of the packet, but also reshapes its spectral distribution. 
Indeed, as time evolves wavenumbers with larger $\Omega_k>0$ are amplified while those with $\Omega_k < 0$ die out. 
As a result, the spectral center of mass, $K(t)=\sum_k k \lvert c_k(t) \rvert^2/\sum_k \lvert c_k(t) \rvert^2$, initially located at $k_0=0$, drifts toward the wavenumbers where the amplification is stronger, the latter being maximal at $k_\ast=\pi/2$. 
This induces a time-dependent group velocity, $v_g(t)=\left.d\varepsilon_k/dk\right\rvert_{K(t)}$ and thus, an acceleration of the wave-packet center of mass in real space.

In addition, for finite width $\sigma_0$, the initial condition in Eq.~\eqref{eq:initial_conditions_mode} weakly excites eigenmodes in the neighborhood of the fastest-amplifying wavenumber $k_\ast=\pi/2$, with amplitudes scaling as $\left\lvert C_{k_\ast}\right\rvert \propto\left\lvert C_0\right\rvert e^{-\sigma_0^2\pi^2/4}$.
Although these contributions are exponentially small at $t=0$, they can become dominant after sufficient time because their amplification rate is larger than that of the initially dominant eigenmodes around $k_0=0$.
As a result, the real space wave-packet may abruptly relocate from the trajectory associated with $k_0=0$ to that associated with $k_\ast=\pi/2$, provided that the two contributions remain sufficiently coherent and follow distinguishable group velocity trajectories in real space.
This abrupt relocation is referred to as non-Hermitian wave jump~\cite{LME2022,HO2025,JMAFS2026}.

Nonlinearity leads the right-hand side (RHS) of Eq.~\eqref{eq:before_secular_form} to be non-trivial, whenever $k_1 - k_2 + k_3 - k = 0$.
Consequently, the complex amplitude $\phi_k$ is no longer conserved but instead evolves through nonlinear interactions among the eigenmodes.
Interestingly, the RHS of Eq.~\eqref{eq:before_secular_form} stands as a vector with entries containing amplification functions $e^{\lambda_{k, \vec{k}} t}$ at rates, 
\begin{equation}
    \lambda_{k, \vec{k}} = \Omega_{k_1}+\Omega_{k_2}+\Omega_{k_3}-\Omega_k, \quad \vec{k} = \left(k_1, k_2, k_3 \right).
\end{equation}
which can be either positive or negative.
Further, these vector components also exhibits oscillating functions, $e^{i\omega_{k, \vec{k}} t}$ with frequencies,
\begin{equation}
    \omega_{k, \vec{k}} = \varepsilon_k+\varepsilon_{k_2}-\varepsilon_{k_1}-\varepsilon_{k_3}.
    \label{eq:resonance_conditions_02}
\end{equation}

When $\lvert \omega_{k, \vec{k}} \rvert > \lvert \lambda_{k, \vec{k}}\rvert$ we expect the $e^{i\omega_{k, \vec{k}}t}$ function to rapidly oscillate compared to the amplification.
It follows that these terms average out over time when looking at their contributions to the variations of the complex amplitude, $\phi_k$.
On the other hand, when $\lvert \omega_{k, \vec{k}} \rvert < \lvert \lambda_{k, \vec{k}}\rvert$ the period of the oscillating function is larger than the amplification time scale.
As such, these terms do not cancel out when performing a time average of their contributions.
The same argument is valid for certain quadruplets $(k, \vec{k})$, whose values $\lvert \omega_{k, \vec{k}} \rvert = 0$.
The two latter are respectively near-secular and secular terms, defining some slow evolution of the complex amplitude $\phi_k$, see also~\cite{C1979,LO2018,LZBKC2019,MCTS2022}.

Averaging over time the RHS of Eq.~\eqref{eq:eqs_mot_nonlinear_NMspace}, only the secular and near-secular terms survive. 
The resulting secular equations take the form
\begin{equation}
    -i\frac{d\phi_k}{dt} = \frac{\alpha}{N} \sum_{k_1,k_2,k_3}\phi_{k_1}\phi_{k_2}^\star\phi_{k_3} e^{\lambda_{k, \vec{k}} t}.
    \label{eq:secular_form}
\end{equation}
In order to get some analytical insights, we approximate the RHS of Eq.~\eqref{eq:secular_form} using the trivial quadruplets $k_1 = k_2 = k_3 = k$.
It follows that the secular form equations,
\begin{equation}
    -i\frac{d\phi_k}{dt} = \frac{\alpha}{N}|\phi_k|^2\phi_k\,e^{2\Omega_k t}.
    \label{eq:self_phase_only}
\end{equation}
can now be easily solved, when considering initial conditions $\phi_k(0)=C_k$.
We find,
\begin{equation}
    \phi_k(t)= C_k e^{i\Gamma_k(t)},~~ \mbox{with}~~ \Gamma_k(t) = 
    \frac{\alpha |C_k|^2}{N}\dfrac{e^{2\Omega_k t}-1}{2\Omega_k},
    \label{eq:nonlinear_phase_shift}
\end{equation}
considering that $\lvert\phi_k(t)\rvert^2 = \lvert C_k \rvert^2$ is a conserved quantity.
Remarkably, the diagonal elements of the coupling tensor in Eq.~\eqref{eq:self_phase_only} lead to a renormalization of the frequencies of the eigenmodes from $E_k$ to $\widehat{E}_k(t) = \widehat{\varepsilon}_k(t) - i\Omega_k$ with $\widehat{\varepsilon}_k(t) = \varepsilon_k + \nu_k(t)$.
The resulting nonlinear frequency shift
\begin{equation}
    \nu_k(t) = \frac{d\Gamma_k}{dt}, ~ \mbox{ gives } ~ 
    \nu_k(t) = \frac{\alpha \lvert C_k \rvert^2}{N}e^{2\Omega_kt}.
\end{equation}
Moreover, since the complex amplitude can be written as $ c_k (t) \propto C_k e^{\Omega_k t} e^{i\widehat{\varepsilon}_k t}$ the shape and phase of these {\it nonlinear eigenmodes} are also preserved.
It follows that they can interact with their linear counterparts.
Crucially, the renormalized frequencies of these nonlinear eigenmodes explicitly depend on time, shifting the excited part of the spectrum along the real axis of the complex plane.



\subsection{\label{subsec:expected_regimes_kspace}Expected dynamical regimes of wave-packet excitations}

Let us now discuss the fate of the initial Gaussian wave-packet in Eq.~\eqref{eq:initial_condition_discrete}.  
In the eigenmode basis, the initial condition reads, $c_k (t=0) = C_0 e^{-\sigma_0^2 k^2}$, within the eigenmode basis, the total norm at $t=0$ gives $\widetilde{S}= \lvert C_0 \rvert^2\sqrt{\pi}/\sigma_0 \sqrt{2}$.
Notably, this initial condition excites a large number of eigenmodes, $\widetilde{L} = N/2\sigma_0 \sqrt{\pi}$ ($N\gg \sigma_0$).
Assuming that these excited eigenmodes have approximately the same initial amplitude, $\widetilde{s}_0=\widetilde{S}/\widetilde{L}$, we express the time evolution of the nonlinear frequency shift,
\begin{equation}
    \nu(t) \sim  \alpha N^{-1} \lvert \widetilde{s}_0\rvert^2 \exp \left(\frac{bt}{2\sigma_0} \right), \quad b= g-1,
    \label{eq:explicit_nonlinear_freq_shift}
\end{equation}
of the excited eigenmodes with $\Omega_k >0$.
It is worth emphasizing that Eq.~\eqref{eq:explicit_nonlinear_freq_shift} is derived using the $\nu_k(t)$ of fastest amplifying initially excited eigenmode, whose wavenumber $k$ is proportional to  $1/\sigma_0$.

This estimate allows for the identification of the expected dynamical regimes of the wave-packet evolution.
As stated previously, the $\nu(t)$ [Eq.~\eqref{eq:explicit_nonlinear_freq_shift}] grows in time and can be compared to the linear frequency scales $d$ [Eq.~\eqref{eq:average_spacing}] and $\Delta$ [Eq.~\eqref{eq:width_frequency}].
Thus, considering small nonlinearities, $\alpha\ll 1$, we find that at short times, the $\nu (t) < d$.
In this case, we see the nonlinear frequency shift remains small such that the eigenmodes are weakly interacting with each other.
It follows the wave-packet evolves perturbatively close to its linear dynamics. 
We therefore refer to this phase as the \textit{nonlinear-skin} regime.

As time evolves, the $\nu(t)$ grows due to wave amplification and eventually reaches values  $\nu(t) \simeq d$ with $\nu(t) < \Delta$.
In this case, the perturbative argument above begins to break down, since nonlinear interactions can no longer be neglected. 
These interactions take the form of resonances between the excited eigenmodes with the unexcited ones and progressively activate a larger fraction of the spectrum. 
As a result, the initial wave-packet is expected to lose its shape. 
We refer to this second phase as the \textit{wave-mixing} regime.

Finally, at sufficiently long times, amplification drives the nonlinear frequency shift to values $\nu (t) > \Delta$.
In this case, some excited eigenmodes are tuned out of resonance with the neighboring ones, suppressing efficient energy exchange between them.
This may lead to the formation of persistent discrete localized structures characteristic of the \textit{self-trapping} regime.
To summarize, the different dynamical regimes of a wave-packet in the nonlinear HN model are as follows:
\begin{align*}
    \nu(t)<d &: \text{nonlinear-skin},\\
    d<\nu(t)<\Delta &: \text{wave-mixing},\\
    \nu(t)>\Delta &: \text{self-trapping}.
\end{align*}
It is worth emphasizing that, for a Gaussian pulse, these criteria are expected to provide only rough estimates of the boundaries between the different dynamical regimes, since $\nu(t)$ can itself only be approximated, see Eqs.~\eqref{eq:secular_form} to~\eqref{eq:explicit_nonlinear_freq_shift}, see also Refs.~\cite{F2010,LBKSF2010}.

\section{\label{sec:analytic_spreading}Estimating the properties of Wave-packet propagation}

We adopt a continuum approximation by identifying the discrete site index with the continuous coordinate, $x=n$, and replacing $y_n(t)$ with $y(x,t)$. 
We further apply the transformation $y\rightarrow ye^{i\varepsilon_0 t}$, where $\varepsilon_0=-(g+1)$; see Appendix~\ref{secapp:continuum}.
This procedure generates the following partial differential equation (PDE) of motion
\begin{equation}
        -i \frac{\partial y}{\partial t} =  -b \frac{\partial y}{\partial x} + D \frac{\partial^2 y}{\partial x^2} + \alpha \lvert y \rvert^2 y,
        \label{eq:pde_of_motion}
\end{equation}
where $b=g-1$ and $D=(g+1)/2$. 
The initial condition for the PDE above has the same form as in Eq.~\eqref{eq:initial_condition_discrete}, with $(n,n_0)\rightarrow(x,X_0)$. 
Equation~\eqref{eq:pde_of_motion} is valid when higher-order dispersive effects can be neglected, for example at long wavelengths and short times~\cite{LW2022,CNX2024,HO2025,JMAFS2026}. 
Further, this PDE provides a more accurate prediction of the above limit than the one reported in Ref.~\cite{LW2022}, see Sec.~\ref{subsec:linear_transport}.
As we will show below, the above represents a perfect setting for isolating the role of nonlinearity, in close analogy with the same approach used to study the effects of dispersion in the HN chain~\cite{HO2025,JMAFS2026}.

\subsection{\label{subsec:linear_transport}Linear limit}
It is useful to first consider the linear theory of wave-packets of the HN model, as it provides a reference for understanding its nonlinear counterpart. 
Starting from the PDE of motion [Eq.~\eqref{eq:pde_of_motion}], the linear limit is obtained, for example by setting $\alpha  = 0$. 
In this limit, the solution of Eq.~\eqref{eq:pde_of_motion} can be obtained explicitly, yielding~\cite{GS2018,LW2022,JMAFS2026}
\begin{equation}
    \begin{split}
        y(x, t) = \dfrac{1}{\sqrt[4]{2\pi\sigma^2 (t)}} \exp \left( \frac{b^2t^2}{4\sigma_0^2} \right) \exp \left[ - \frac{\left(x - X(t)\right)^2}{4\sigma^2 (t)} \right],
        \label{eq:solution_wavepacket_linear}
    \end{split}
\end{equation}
where $\sqrt[4]{2\pi \sigma^2 (t)}$ plays the role of a renormalization factor.
It follows that,
\begin{equation}
    S(t) = \exp \left( \frac{b^2t^2}{2\sigma_0^2} \right),
    \label{eq:linear_amplification}
\end{equation}
gives the time evolution of the norm of the wave-packet.
Furthermore, the first and second moments of position of its norm density,
\begin{equation}
    X(t) = X_0 + \frac{bD t^2}{\sigma_0^2}, \quad \sigma^2 (t) = \sigma_0^2 + \dfrac{D^2 t^2}{\sigma_0^2},
    \label{eq:width_and_linear}
\end{equation}
describe the time evolution of the wave-packet center of mass and width in real space.

Consequently, the dynamics of a bulk wave-packet in the weakly dispersive HN model is characterized by a monotonic increase in its spatial extent.
The latter is caused solely by dispersion, whose strength is governed by the parameter $D$, calculated from the real part of the eigenfrequency spectrum.
Moreover, this wave-packet center of mass also undergoes constant acceleration, in the direction of amplification (NHSE) driven by the combined effects of nonreciprocity and dispersion, through the parameters $b$ and $D$.
On the other hand, its norm, $S$, grows super-exponentially in time, at a rate proportional to the squared nonreciprocal parameter $b$.

\subsection{\label{subsec:nonlinear_acceleration} Weak nonlinear limit of the nonlinear-skin regime}

In this regime, nonlinear interactions remain weak. 
To approximate the time evolution of the initial Gaussian wave-packet in the nonlinear HN model, we therefore adopt the Gaussian ansatz~\cite{DAL1991,M1995,CBG1996,PMCLZ1997}
\begin{equation}
    \begin{split}
            &y(x,  t) =  A(t) \exp \left[-\frac{\left(x - X(t) \right)^2}{4\sigma ^2 (t)}\right] \times \\
            &\exp \left[i \Phi(t) + i\rho(t)\left(x - X(t) \right) + i\eta(t)\left(x - X(t) \right)^2\right]
    \end{split}
    \label{eq:ansatz_nonlinear}
\end{equation}
where the time-dependent variational parameters $A$, $\sigma$, $X$, $\Phi$, $\rho$, and $\eta$ are to be determined.
This trial solution is fully compatible with the linear limit of the HN lattice, like Eq.~\eqref{eq:solution_wavepacket_linear}.
Further, to find the variables of the nonlinear solution [Eq.~\eqref{eq:ansatz_nonlinear}], we employ the collective coordinate approach, considering  that the parameters of the ansatz are independent~\cite{CBG1996}.

Substituting this ansatz into the HN PDE of motion and after straightforward although tedious simplifications (see Appendix~\ref{secapp:nonlinear_waves}), we find the governing equations of each of the variational parameters,
\begin{eqnarray} 
    \label{eq:var_param_1}
    \dot{A} &=& b\rho A - 2D\eta A, \\ 
    \label{eq:var_param_2}
    \dot{X} &=& 2D \rho + 4b\eta \sigma^2, \\ 
    \label{eq:var_param_3}
    \dot{\sigma} &=&4D\eta \sigma, \\ 
    \label{eq:var_param_4}
    \dot{\rho} &=& \frac{b}{2\sigma^2} + 8 b\eta^2 \sigma^2, \\ \label{eq:var_param_5}
    \dot{\eta} &=& \frac{D}{4\sigma^4} - 4D\eta^2 - \frac{\alpha A^2\sigma}{4\sqrt{2}\sigma^2}.
\end{eqnarray}
Consequently, the set of Eqs.~\eqref{eq:var_param_1}--\eqref{eq:var_param_5} completely characterizes the evolution of the Gaussian wave-packet in the HN model with nonlinear self-interactions.

Let us now discuss in detail the evolution equations for the parameters above.
Assuming that at all times, the norm of the wave-packet,
\begin{equation}
    S = \sqrt{2\pi} A^2 \sigma,
    \label{eq:total_amplitude_perturbation}
\end{equation}
is known, Eqs.~\eqref{eq:var_param_1} and~\eqref{eq:var_param_3} capture its non-conservation.
Indeed, the $\dot{S}=2b\rho S$ is always nontrivial in presence of nonreciprocity, $g>1$.
In particular, at short times within the nonlinear-skin regime, it is safe to consider that the weak nonlinearity has little effects on the already super-exponential amplification found in the linear limit.
It follows that Eq.~\eqref{eq:linear_amplification}, captures the nonlinear wave-packet amplification [Eq.~\eqref{eq:total_amplitude_perturbation}] with good accuracy.
We will check in Sec.~\ref{sec:numerical_results} the validity of this approximation.

Moving on, we eliminate the expressions of $\rho$ and $\eta$ in Eq.~\eqref{eq:var_param_3}.
We obtain that the width satisfies the following second-order equation,
\begin{equation}
    \ddot{\sigma} =\dfrac{\displaystyle D^2}{\displaystyle \sigma^3} - \dfrac{\displaystyle  \alpha D S}{\displaystyle 2\sqrt{\pi}}\dfrac{\displaystyle 1}{\displaystyle \sigma^2}
    .
    \label{eq:ode_nonlinear_width_acceleration_01}
\end{equation} 
As such nonlinearity couples the amplification, $S$, to the observable and parameter of dispersion, $\sigma$ and $D$ respectively.
Interestingly, the governing equation of the nonlinear width, is generated by a time-dependent Hamiltonian dynamics~\cite{GPS2002}, $\ddot{\sigma}=- \partial H_\sigma/\partial \sigma$ with 
\begin{equation}
    H_\sigma = K_\sigma + U_\sigma, ~ K_\sigma = \frac{\dot{\sigma}^2}{2}, ~U_\sigma = \frac{D^2}{2\sigma^2} - \frac{\alpha D S}{2\sqrt{\pi}\sigma}.
    \label{eq:hamiltonian_width}
\end{equation}
Here $K_\sigma$ and $U_\sigma$ are the effective  kinetic and potential functions.

\begin{figure}[!tb]
    \centering
    \includegraphics[width=\columnwidth]{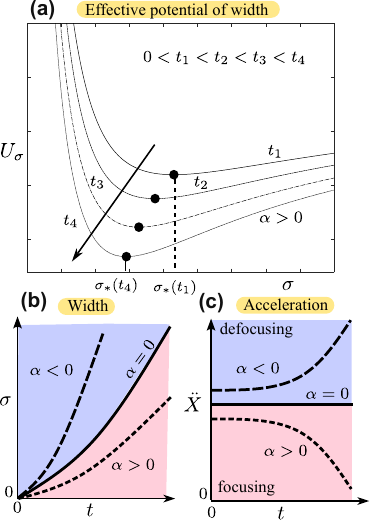}
    \caption{
    {\it Analytical estimates within the nonlinear-skin regime.} 
    (a) Schematic of the effective potential $U_\sigma$ associated with the Hamiltonian formulation of the equation governing the nonlinear wave-packet width for focusing nonlinearity, $\alpha>0$.
    The corresponding instantaneous fixed point $\sigma_\ast$ shifts toward smaller values as time increases.
    (b) Time dependence of the wave-packet width $\sigma$ in the linear and nonlinear cases.
    Focusing nonlinearity causes $\sigma$ to grow more slowly than in the linear case (dotted curve in the red region), whereas defocusing nonlinearity enhances its growth (dashed curve in the blue region).
    (c) Same as in panel (b), but for the wave-packet acceleration $\ddot{X}$.
    Focusing nonlinearity reduces the magnitude of $\ddot{X}$ and causes it to decrease with time (dotted curve in the red region), whereas defocusing nonlinearity enhances the acceleration and causes it to increase (dashed curve in the blue region).
    See text for details.
    }
    \label{fig:analytical_results}
\end{figure}

Equation~\eqref{eq:ode_nonlinear_width_acceleration_01} cannot be easily integrated, except in the linear limit where we recover Eq.~\eqref{eq:width_and_linear} when considering $\sigma(t=0)=\sigma_0$.
In order to get insight within the nonlinear phase, the analysis of the equilibrium point
\begin{equation}
    \left(\sigma_\ast = \frac{2D\sqrt{\pi}}{\alpha S}, \quad \dot{\sigma}_\ast = 0\right),
    \label{eq:fixed_point_width}
\end{equation}
of Eq.~\eqref{eq:ode_nonlinear_width_acceleration_01} is much more illuminating.
This fixed point corresponds to a stable minimum $U_{\sigma}(\sigma_\ast) = -\alpha^2S^2/8\pi$, with $U_{\sigma}^{\prime \prime} (\sigma_\ast) = D^2/\sigma_\ast^4>0$.
Importantly, the position and Hamiltonian of this fixed point exponentially decay in time. 
This is illustrated in Fig.~\ref{fig:analytical_results}(a), where the potential is plotted at different times for $g=1.25$, $\sigma_0=8$, and $\alpha=6$ (chosen for clear visualization). 
It follows that solutions $\sigma(t)$ of Eq.~\eqref{eq:ode_nonlinear_width_acceleration_01} that remain trapped within the potential well around $\sigma_\ast$ are expected to exhibit a similar trend, Fig.~\ref{fig:analytical_results}(a).

Furthermore, taking a small perturbation about $\sigma_\ast$, evaluated at $t=0$, as the initial condition for the width equation, we find that the corresponding linear evolution satisfies $\ddot{\sigma}\approx\alpha^3S^3/(8D\pi^{3/2})$, whereas $\ddot{\sigma}\approx0$ in the nonlinear evolution. 
Thus, at early times, the rate of growth of the wave-packet width is larger in the linear case than in its focusing nonlinear counterpart where $\alpha>0$. 
As the width increases monotonically with time, the nonlinear width will remain smaller than its linear counterpart throughout the time evolution, as shown in Fig.~\ref{fig:analytical_results}(b).

On the other hand, for defocusing nonlinearities with $\alpha<0$, the effective Hamiltonian in Eq.~\eqref{eq:hamiltonian_width} has no physically meaningful fixed point, since the corresponding value of $\sigma_\ast$ is negative. In this case, $\ddot{\sigma}$ [Eq.~\eqref{eq:ode_nonlinear_width_acceleration_01}] remains positive and increases more rapidly with time than in the linear limit. We therefore expect the wave-packet width to exceed its linear counterpart throughout the evolution, as shown in Fig.~\ref{fig:analytical_results}(b). These results are consistent with the general effects of on-site nonlinearities in the HN lattice~\cite{MCKA2024}.

Turning now to the motion of the wave-packet center of mass, we eliminate $\rho$ and $\eta$ from Eq.~\eqref{eq:var_param_2} and obtain the following second-order equation governing the wave-packet acceleration, 
\begin{equation}
    \ddot{X} =  \dfrac{\displaystyle 2b D}{\displaystyle \sigma^2} + \dfrac{\displaystyle 2b}{\displaystyle D}\dot{\sigma}^2 - \dfrac{\displaystyle \alpha b S}{\displaystyle 2\sqrt{\pi}}\dfrac{\displaystyle 1}{\displaystyle \sigma}.
    \label{eq:ode_nonlinear_width_acceleration_02}
\end{equation} 
For $\alpha=0$, Eq.~\eqref{eq:ode_nonlinear_width_acceleration_02} gives
$\ddot{X}=2bD/\sigma^2+2b\dot{\sigma}^2/D$, which reduces to $\ddot{X}=2Db/\sigma_0^2$, in agreement with the linear theory developed in Sec.~\ref{subsec:linear_transport}. 
For $\alpha \neq 0$, however, nonlinearity couples nonreciprocity, amplification, and dispersion, while the resulting wave-packet acceleration remains fundamentally of non-Hermitian origin.
Remarkably, already at $t=0$, nonlinearity modifies the magnitude of the acceleration: focusing nonlinearities reduce it, whereas defocusing nonlinearities enhance it.

To meaningfully analyze the time dependence of this nonlinear acceleration, it is
useful to compute its rate of change
\begin{equation}
    \frac{d}{dt} \ddot{X} = - \frac{\alpha b S}{2\sigma \sqrt{\pi}} \left(\frac{\dot{S}}{S} + 3 \frac{\dot{\sigma}}{\sigma} \right).
    \label{eq:deriv_acceleration}
\end{equation}
This expression is always trivial in the linear case, such that the acceleration is constant as expected.
In case $\alpha \neq 0$, we obtain a time-dependent acceleration $\ddot{X}$.

More specifically, keeping in mind that the nonlinear-skin regime is a perturbation of the linear limit, we expect the amplification and the broadening of the pulse to exhibit a super-exponential and a polynomial-like dependence against time.
As a result, the sum of the relative rates $\dot{S}/S + 3\dot{\sigma}/\sigma$ is always positive irrespective of the nonlinear parameter.
This can be clearly illustrated using Eq.~\eqref{eq:width_and_linear}.
In this scenario we find $\dot{S}/S \sim  t$ and $\dot{\sigma}/\sigma \rightarrow 0$, leading to $\dot{S}/S + 3\dot{\sigma}/\sigma \sim t$ with $t>0$.
It follows that values of the RHS of Eq.~\eqref{eq:deriv_acceleration} solely depend on the sign of the nonlinear coefficient, $\alpha$.
Thus focusing nonlinearities, characterized by $\alpha >0$, lead to $d \ddot{X}/dt <0$.
That is to say, the acceleration decreases in time, Fig.~\ref{fig:analytical_results}(c).
On the other hand, for defocusing nonlinearities where $\alpha <0$, we obtain $d\ddot{X}/dt >0$, resulting in an acceleration that grows as time increases, Fig.~\ref{fig:analytical_results}(c).

We verified the arguments above by numerically integrating Eqs.~\eqref{eq:ode_nonlinear_width_acceleration_01} and~\eqref{eq:ode_nonlinear_width_acceleration_02} subject to the initial conditions
$(\sigma,X,\dot{\sigma},\dot{X})=(\sigma_0,0,0,0)$. 
This procedure leads to high-accuracy predictions for the time evolution of the wave-packet width and center of mass across the range of system control parameters.

\subsection{\label{subsec:selftrapping}Ballistic spreading of the wave-packet in the self-trapping regime}
We proceed by analyzing the self-trapping regime in more detail.
A trapped bulk wave can naturally be divided into two distinct spatial regions: a central core and tails.
The sites within the central core are characterized by large amplitudes.
Consequently, in this region, the nonlinear terms dominate the dynamics, leading to the following equation of motion and solutions
\begin{equation}
    \begin{split}
            -i\frac{dy_n}{dt} \sim \nu_n y_n, \quad \mbox{and} \quad 
            y_n(t) \sim Y_n e^{i\nu_n(t_0) \tau}
    \end{split}
    \label{eq:selftrapping_core}
\end{equation}
where $\tau = t - t_0$, $\nu_n = \alpha \lvert y_n \rvert^2$.
We solved equations of motion in Eq.~\eqref{eq:selftrapping_core} using the initial condition, $y_n(t_0)=Y_n\delta_{n,m}$ with $t_0>T_{\mathrm{trap}}$.
We recall that $\delta_{n,m} = 1$ if $n=m$ otherwise $\delta_{n,m}=0$.
Consequently within the trapped region, amplitudes remain large and rapidly oscillate in time, with $\nu_n(t)\gg 1$.

On the other hand, an oscillator within the tails of the wave-packet satisfies $\lvert y_n (t) \rvert^2 \to 0$.
Its dynamics is therefore well approximated by the linear equations~\cite{KLM2014,E2022,SH2025},
\begin{equation}
    \begin{split}
        -i\frac{dy_n}{dt} \sim \sum_r h_{n,r}y_r, \mbox{ and } 
        y_n(t) \sim Y_n\left(iG\right)^{\ell} J_{|\ell|} \left(\theta\right),
    \end{split}
    \label{eq:selftrapping_tails}
\end{equation}
with, $\ell = n-m$, $G=\sqrt{g}$, $\theta = 2\tau G$ and $\tau=t-t_0$.
Note that $h_{n,r}=1$ and $h_{n,r}=g$ for $r=n+1$ and $r=n-1$, respectively, and $h_{n,r}=0$ otherwise.
In addition, $J_{|\ell|}$ denotes the Bessel function of the first kind~\cite{AS1964} and we use the same initial condition as in Eq.~\eqref{eq:selftrapping_core}.
The solution in Eq.~\eqref{eq:selftrapping_tails} results in~\cite{KLM2014,SH2025}
\begin{equation}
    \sigma^2 \sim t^2
    \label{eq:spreading_law_core}
\end{equation}
corresponding to ballistic propagation of the wave-packet's tails in the direction favored by the NHSE.
Behind the propagating front, the excited oscillators undergo nonreciprocal wave-mixing and amplification, which eventually promotes the formation of new trapped states that accumulate within the wave-packet core.

\section{\label{sec:regime_transition}Characteristic time scales and absence of non-Hermitian jump}

\subsection{\label{subsec:time_scales}Characteristic time scales of nonlinear wave-packet dynamics}

We now determine the dynamical regime boundary times using the framework of Sec.~\ref{subsec:expected_regimes_kspace}. 
In doing so, we consider the initial pulse of Eq.~\eqref{eq:initial_condition_discrete}, spanning a length $ L = 2\sigma_0\sqrt{\pi}$.
Assuming that the norm $S(0)$ is approximately uniformly distributed over these $L$ excited sites, the initial average nonlinear frequency shift is $\nu(0) = \alpha S(0)/L$.
In addition, its time evolution is approximately given by
$\nu(t)\sim\alpha(\sqrt{2}\sigma_0)^{-1}\exp\left[b^2t^2/(2\sigma_0^2)\right]$.
It is worth emphasizing that this expression provides a more representative estimate of the nonlinear frequency shift than Eq.~\eqref{eq:explicit_nonlinear_freq_shift}, since the real space pulse dynamics effectively averages over the excited eigenmodes.
Consequently, in the parameter space defined by the nonlinear coefficient $\alpha$ and time $t$, two characteristic time scales naturally emerge. 
The first one,
\begin{equation}
    T_{\mathrm{mix}}^2 \sim \frac{2\sigma_0^2}{b^2}
    \ln\left[\frac{4\sqrt{2}\sigma_0D}{\alpha N}\right],
    \label{eq:tmix}
\end{equation}
calculated from $\nu(t) = d$, marks the boundary between the nonlinear-skin and wave-mixing regimes while the second one,
\begin{equation}
    T_{\mathrm{trap}}^2 \sim \frac{2\sigma_0^2}{b^2}
    \ln\left[\frac{4\sqrt{2}\sigma_0D}{\alpha}\right],
    \label{eq:ttrap}
\end{equation}
defines the boundary between the wave-mixing and self-trapping regimes.
The latter being estimated through $\nu(t) = \Delta$.
The dynamical regimes can therefore be equivalently expressed in terms of these times as
\begin{align*}
    t < T_{\mathrm{mix}} &: \text{nonlinear-skin},\\
    T_{\mathrm{mix}} < t < T_{\mathrm{trap}} &: \text{wave-mixing},\\
    t > T_{\mathrm{trap}} &: \text{self-trapping}.
\end{align*}

Figure~\ref{fig:schematic}(d) illustrates these dynamical regimes for an initial Gaussian pulse in the nonlinear HN lattice. 
The lower and upper black curves correspond to $T_{\mathrm{mix}}$ and $T_{\mathrm{trap}}$ respectively plotted for $g=1.25$, $\sigma_0 = 8$, $N = 100$, $S(0)=1$ and $\alpha$ varying between $0.01$ to $100$.
Consequently, in the region below $T_{\mathrm{mix}}$, we expect the wave-packet to follow the predictions of Sec.~\ref{subsec:nonlinear_acceleration} within the nonlinear-skin regimes.
Further, the region between $T_{\mathrm{mix}}$ and $T_{\mathrm{trap}}$, denotes the wave-mixing regimes featuring energy exchange between eigenmodes, and thus a loss of coherence of the nonlinear wave-packet.
Beyond $T_{\mathrm{trap}}$, the system enters a self-trapping regime characterized by long-lived coherent structures wandering within the excited region of the lattice.

It is worth commenting on the validity of the explicit analytical expressions of $T_{\mathrm{mix}}$ and $T_{\mathrm{trap}}$.
Indeed, their analytical derivations are based on the simplified secular form [Eq.~\eqref{eq:self_phase_only}] which considers only the diagonal terms of the eigenmodes' coupling tensor in Eq.~\eqref{eq:secular_form}.
A different outcome is expected due to the contribution of the off-diagonal terms of the coupling tensor above (non-trivial wave-mixing).
Further, the nonsecular terms, neglected within the secular form equations, also contribute to energy exchanges between eigenmodes in finite times.
On the other hand, the nonlinear frequency shift induced by the wave-packet can only be approximated due to the large number of initially excited eigenmodes and thier non-trivial dynamics. 
It follows that the boundaries defined by these time scales are qualitative and not sharp.
They primarily indicate the dependence of the three dynamical regimes identified above on the control parameters.

\subsection{\label{subsec:jump}Absence of non-Hermitian jump for nonlinear wave-packets}

Let us now estimate the time scale at which the non-Hermitian wave jump occurs.
As discussed in Sec.~\ref{subsec:secular_form}, this jump results from the competition between the initially dominant wavenumber $k_0=0$ and the maximally amplified wavenumber $k_\ast=\pi/2$.
To describe this competition in real space, we use the continuum approximation in the discrete HN model to describe relatively wide localized pulses centered around each of these wavenumbers; see Appendix~\ref{secapp:continuum}.
The packet centered at $k_0=0$ is initialized by Eq.~\eqref{eq:initial_condition_discrete}, while the one at $k_\ast=\pi/2$ is initialized with $y_n^{(\ast)}(0)=e^{-\sigma_0^2\pi^2/4} e^{ik_\ast n}y_n(0)$,  where the factor $e^{-\sigma_0^2\pi^2/4}$ accounts for its exponentially small spectral weight.
The $k_0=0$ continuum dynamics is derived explicitly in Sec.~\ref{subsec:linear_transport}.
Estimating the non-Hermitian wave jump characteristic time as the instant at which the norms of these two pulses become comparable, we obtain
\begin{equation}
    T_{\mathrm{jump}} \sim
    \frac{\pi^2\sigma_0^2}{4b}.
    \label{eq:time_jump}
\end{equation}
Note that a similar result is obtained by working directly in eigenmode space.
As found in previous studies~\cite{HO2025,JMAFS2026} the time at which the non-Hermitian jump happens increases quadratically with the initial packet width, $\sigma_0$, and is inversely proportional to the nonreciprocal strength, $b=g-1$.

So is there a non-Hermitian wave jump in the nonlinear HN lattice?
The short answer is no.
To understand why, we compare the characteristic time scale of the non-Hermitian wave jump, $ T_{\mathrm{jump}}$ [Eq.~\eqref{eq:time_jump}] with the nonlinear time scale introduced above, $T_{\mathrm{mix}}$ [Eq.~\eqref{eq:tmix}].
Because the nonlinear frequency shift grows exponentially in time, these time scales remain relatively small, as reflected by their logarithmic dependence on the nonlinear coefficient, $\alpha$.
As a result, in the presence of even a small nonlinearity, the system typically reaches the mixing threshold $T_{\mathrm{mix}}$ before the wave-packet can undergo the coherent spectral rearrangement responsible for the non-Hermitian wave jump.

Quantitatively, according to Eqs.~\eqref{eq:tmix} and \eqref{eq:time_jump}, the condition $T_{\mathrm{jump}}\lesssim T_{\mathrm{mix}}$  
yields the nonlinear threshold
\begin{equation}
    \alpha_{jump} N \lesssim 4\sqrt{2}\sigma_0D \exp\left(-\frac{\pi^4\sigma_0^2}{32}\right),
    \label{eq:condition_jump}
\end{equation}
for which the non-Hermitian wave jump may be observed.
Considering a wave-packet of width $\sigma_0=8$, gives
$\alpha_{\mathrm{jump}} N\lesssim 10^{-83}$, leading to extremely weak nonlinearities.
More generally, since the condition $\sigma_0>4\sqrt{2}/\pi^2\simeq0.57$ is satisfied for any initial wave-packets extending over a few lattice sites and the magnitude of the exponent in Eq.~\eqref{eq:condition_jump} is quadratic in $\sigma_0$, the threshold $\alpha_{\mathrm{jump}}N$ remains exponentially small. 
Consequently, even a weak nonlinearity can exceed this threshold and prevent the non-Hermitian jump from occurring.

\begin{figure}[!tb]
    \centering
    \includegraphics[width=\columnwidth]{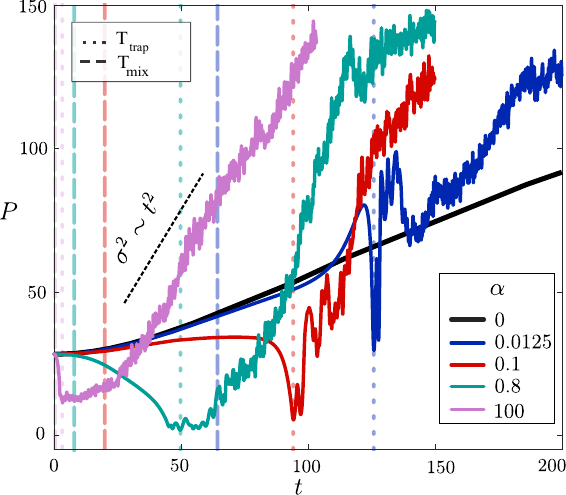}
    \caption{{\it Numerical results}.
                Dependence of the participation number, $P$, against time, $t$, for a HN lattice with $N=512$ and $g=1.25$.
                As the initial condition, we use the Gaussian wave-packet of Eq.~\eqref{eq:initial_condition_discrete} with $\sigma_0=8$, $X_0 = 0$, and $k_0 = 0$.
                The curves correspond to cases with $\alpha=0$ (black), $\alpha=0.0125$ (blue), $\alpha=0.1$ (red), $\alpha=0.8$ (cyan), and $\alpha=100$ (magenta).
                Color-matched dashed lines mark the times $T_{\mathrm{mix}}$ at which the nonlinear curves first depart from the linear-like behavior represented by the black curve.
                Similarly, color-matched dotted lines mark the times $T_{\mathrm{trap}}$ at which $P(t)$ reaches its lowest local minimum after $t=0$.
                The black dotted line guides the eye for $P(t)\sim t$, see text for details.
    }
    \label{fig:participation_ratio_nonrecip_weak_all}
\end{figure}

\section{\label{sec:numerical_results}Numerical results}

\subsection{\label{subsec:num_spreading}Characteristics of nonlinear wave-packet dynamics}
\begin{figure}[!tb]
    \centering
    \includegraphics[width=\columnwidth]{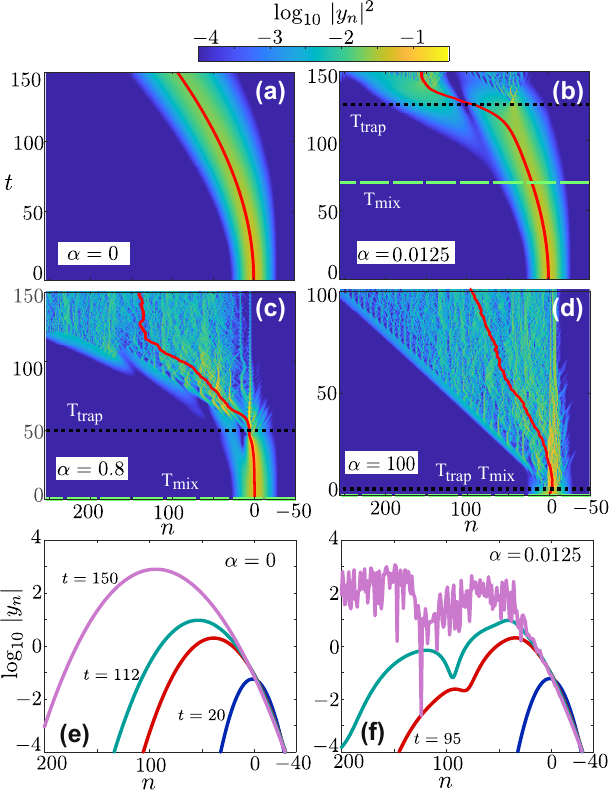}
    \caption{{\it Numerical results.}
        (a)-(d) Spatio-temporal evolution of the norm density $\lvert y_n(t)\rvert^2$ in a lattice with $N=512$ sites, OBCs and $g=1.25$.
        The $\lvert y_n(t)\rvert^2$ is normalized at each time step for clarity.
        The system is initialized with a Gaussian wave-packet [Eq.~\eqref{eq:initial_condition_discrete}] setting $\sigma_0=8$, $n_0=0$, and $k_0=0$, see also Fig.~\ref{fig:participation_ratio_nonrecip_weak_all}.
        Four representative cases are shown: (a) $\alpha=0$, (b) $\alpha=0.0125$, (c) $\alpha=0.8$, and (d) $\alpha=100$.
        In all panels, points are colored according to the magnitude of the norm density according to the color scale at the top. 
        The red curves in (a)-(d) show the dependence against time of the mean position of the norm density [Eq.~\eqref{eq:center_of_mass_numerical}]. 
        In addition, the green dashed and dotted horizontal lines mark the characteristic times $T_{\mathrm{mix}}$ and $T_{\mathrm{trap}}$, which separate the nonlinear-skin, wave-mixing, and self-trapping regimes.
        Panels (e) and (f) show snapshots of the spatial profiles at times $t=20$ (blue), $t=95$ (red), $t=112$ (cyan), and $t=150$ (magenta) for (e) $\alpha=0$ and (f) $\alpha=0.0125$. 
    }
    \label{fig:heatmaps_real_space_01a}
\end{figure}

In this section, we present the computational methods and numerical results used to verify the existence of the dynamical regimes identified above.
As physical observables, we monitor the norm density profile $s_n(t)=\lvert y_n(t)\rvert^2$, along with the total norm and participation number,
\begin{equation}
S(t)=\sum_{n=1}^N s_n(t),
~ \mbox{ and } ~
P^{-1}(t)=\left[S(t)\right]^{-2}\sum_{n=1}^Ns_n^2(t).
\label{eq:part_ratio_def}
\end{equation}
The participation number quantifies the spatial extent of the norm density.
In particular, $P(t)\propto \sigma(t)$ for a localized pulse with a single-site pulse leading $P=1$, whereas a uniformly extended one gives $P=N$.
We also measure the mean position
\begin{equation}
   X(t) = \left[ S(t)\right]^{-1}\sum_{n=1}^N n s_n (t),
   \label{eq:center_of_mass_numerical}
\end{equation}
of norm density profile.
For a compact and localized wave-packet, it coincides with the center of mass.

Based on the theoretical predictions derived above, these observables allow us to distinguish the three dynamical regimes.
In the nonlinear-skin regime, the wave-packet remains coherent, and $P(t)$ increases monotonically, exhibiting behavior similar to that found in the linear limit; see Sec.~\ref{subsec:nonlinear_acceleration}.
The onset of the wave-mixing regime is marked by the first noticeable deviation of $P(t)$ from this behavior, signaling the emergence of significant interactions among the eigenmodes and the breakdown of the nonlinear perturbative description underlying the nonlinear-skin regime.
The time at which this deviation occurs defines the boundary between the nonlinear-skin and wave-mixing regimes, $T_{\mathrm{mix}}$.

At later times, the rapid exponential growth of the total norm $S(t)$ drives the formation of a large-amplitude, strongly localized pulse from the initial one.
Consequently, after its initial growth, $P(t)$ is expected to decrease toward values of order $\sigma_0$ or smaller, indicating the onset of self-trapping.
This localization stage is eventually followed by the ballistic spreading of the wave-packet, $P(t)\sim t$.
Accordingly, the boundary between the wave-mixing and self-trapping regimes, $T_{\mathrm{trap}}$ can be identified as the time at which $P(t)$ reaches its lowest local minimum away from $t=0$.
The mean position provides an additional diagnostic.
It evolves smoothly during the coherent propagation characteristic of the nonlinear-skin regime, whereas it may exhibit discontinuity-like variations and fluctuations once the dynamics enters the wave-mixing and self-trapping regimes.

In our computations, the lattice size is fixed to $N=512$ sites to mimic the thermodynamic limit and set OBCs at both ends. 
The equations of motion [Eq.~\eqref{eq:equation_motion_discrete}] are integrated using a Runge–Kutta scheme of order $8$ based on the Dormand–Prince method, $\mbox{DOP}853$~\cite{HNW1993,DMMS2019,freelyDOP853}. 
A time step of $0.005$, a one-step tolerance of $10^{-11}$, and the use of the multi-precision software \mbox{Advanpix}~\cite{advanpix} ensure high computational accuracy.

\begin{figure}[!tb]
    \centering
    \includegraphics[width=\columnwidth]{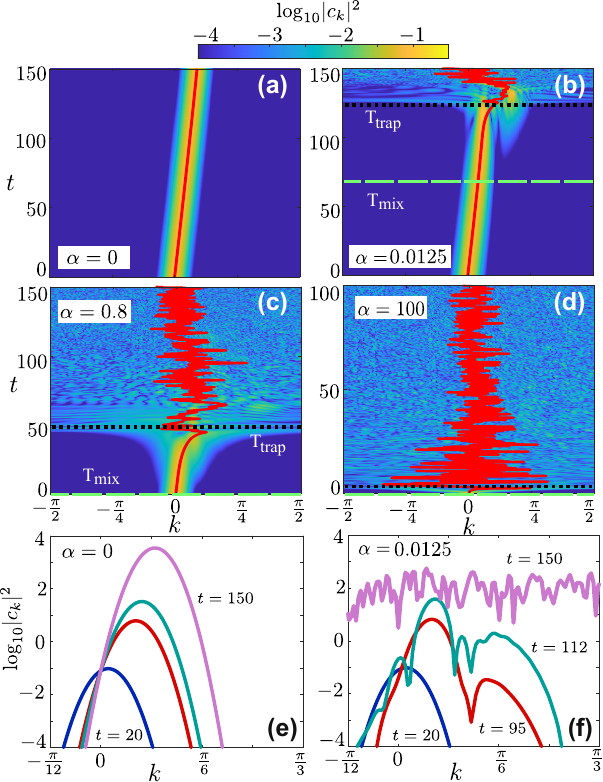}
    \caption{{\it Numerical results.}
        (a-d) Spatio-temporal evolution of the norm density profiles $\lvert c_k (t) \rvert^2$ in the eigenmode variables, $c_k$ 
        (see text for details) using the same results as in Fig.~\ref{fig:heatmaps_real_space_01a}.
        The $\lvert c_k (t) \rvert^2$ is renormalized at each time step for clarity.
        Panels (a)–(d) correspond to the cases with (a) $\alpha=0$, (b) $\alpha=0.0125$, (c) $\alpha=0.8$, and (d) $\alpha=100$. 
        Each point in panels (a)-(d) is colored according to the magnitude of the norm density, as indicated by the color scale at the top.
        We only depict $k\in [-\pi/2, \pi/2]$ for clarity.
        In addition, the red curves in (a)-(d) depict the time dependence of the mean position of the norm density. 
        The green dashed and dotted horizontal lines mark the characteristic times $T_{\mathrm{mix}}$ and $T_{\mathrm{trap}}$, see also Figs.~\ref{fig:heatmaps_real_space_01a}(a)-(d).
        (e)-(f) Snapshots of the norm density profiles at times $t=20$ (blue), $t=95$ (red), $t=112$ (cyan), and $t=150$ (magenta) for (e) $\alpha=0$ and (f) $\alpha=0.0125$. 
    }
    \label{fig:heatmaps_pbc_space_01a}
\end{figure}

Representative examples of the evolution of an initial Gaussian wave-packet of width $\sigma_0=8$, in a weakly nonreciprocal HN chain with $g=1.25$ are shown in Fig.~\ref{fig:participation_ratio_nonrecip_weak_all}, Fig.~\ref{fig:heatmaps_real_space_01a} and Fig.~\ref{fig:heatmaps_pbc_space_01a}. 
Figure~\ref{fig:participation_ratio_nonrecip_weak_all} displays the time dependence of the participation number, $P$, for the linear case, $\alpha = 0$ (black), as well as representative nonlinear cases with $\alpha=0.0125$ (blue), $\alpha =0.1$ (red), $\alpha=0.8$ (cyan), and $\alpha = 100$ (magenta).
The corresponding spatio-temporal evolution of the normalized norm density profiles are shown in Fig.~\ref{fig:heatmaps_real_space_01a} and Fig.~\ref{fig:heatmaps_pbc_space_01a} in the real and eigenmode spaces respectively for (a)  $\alpha = 0$, (b) $\alpha=0.0125$, (c) $\alpha=0.8$, and (d) $\alpha = 100$.

In the linear limit, $P$ grows monotonically with time $t$ [black curve in Fig.~\ref{fig:participation_ratio_nonrecip_weak_all}]. 
This behavior reflects the continuous broadening of the wave-packet due to dispersion, as shown in Fig.~\ref{fig:heatmaps_real_space_01a}(a), and is fully consistent with the linear theory of Sec.~\ref{subsec:linear_transport}. 
Snapshots of the amplitude $\lvert y_n\rvert$ taken at times $t=20$ (blue), $t=95$ (red), $t=112$ (cyan), and $t=150$ (magenta) further show a Gaussian envelope that preserves its overall shape while its spatial extent increases with time. 
Furthermore, its amplitude grows exponentially in time, while its center of mass (red curve) moves toward the boundary favored by the NHSE, see Fig.~\ref{fig:heatmaps_real_space_01a}(e).

In the presence of nonlinearity, the linear picture changes drastically. 
Indeed, for all representative nonlinear cases, $P$ exhibits an overall growth with time, along with pronounced non-monotonic variations even for small nonlinear coefficients, $\alpha=0.0125$ and $0.1$, as illustrated by the blue and red curves in Fig.~\ref{fig:participation_ratio_nonrecip_weak_all}. 
Throughout, we distinguish three different dynamical regimes, which roughly correspond to the regions predicted in Sec.~\ref{sec:regime_transition}. 
The first stage extends from $t=0$ to the times indicated by the dashed horizontal lines. 
Each $P(t)$ curve and its corresponding line markers are shown using a similar color code.
Thus we see that during this stage the participation number grows monotonically and closely follows a linear-like dynamics (black curve in Fig.~\ref{fig:participation_ratio_nonrecip_weak_all}). 
This behavior is clearly visible in the spatio-temporal evolution of the norm density for $\alpha=0.0125$, in Fig.~\ref{fig:heatmaps_real_space_01a}(b), and in the corresponding amplitude snapshots in Fig.~\ref{fig:heatmaps_real_space_01a}(f).
For instance, at $t=20$ within this stage, the blue curve in Fig.~\ref{fig:heatmaps_real_space_01a}(f) shows a wave-packet remaining coherent and exhibiting small amplitudes, with $\max_n \lvert y_n \rvert \approx 0.01$.
Therefore, in this stage the effects of the asymmetric couplings still dominate the dynamics, hence mapping the nonlinear-skin regime of Fig.~\ref{fig:schematic}(c). 

The second stage spans between the times represented by the dashed and dotted horizontal lines in Fig.~\ref{fig:participation_ratio_nonrecip_weak_all}.
There $P$ displays a divergence from a monotonic increase accompanied by a change of concavity (dashed horizontal lines), followed by a slow down of growth.
This slow down is then followed by a decay toward a minimal value of $P$ near the limit marked by the dotted horizontal lines.
This tendency is clearly visible for $\alpha =0.1$ (red), $0.8$ (cyan) and $100$ (magenta) in Fig.~\ref{fig:participation_ratio_nonrecip_weak_all}.
Examining the norm density evolution, we observe that secondary wave-packets separate from the main pulse during this stage. 
This feature is particularly evident for $\alpha=0.0125$ and $\alpha=0.8$; see Figs.~\ref{fig:heatmaps_real_space_01a}(b)–(c). 
The shedding of these secondary envelopes from the main pulse is accompanied by the narrowing and amplification of the latter.
Consequently, the number of strongly excited sites decreases relative to the weakly nonlinear dynamics of the nonlinear-skin regime, causing the growth of the participation number to slow down and eventually reverse into a decay near the dotted horizontal lines in Fig.~\ref{fig:participation_ratio_nonrecip_weak_all}.

Nevertheless, for weak nonlinearities, the secondary wave-packets have sufficient time to grow substantially in amplitude as depicted by Fig.~\ref{fig:heatmaps_real_space_01a}(b) and the cyan curves in Fig.~\ref{fig:heatmaps_real_space_01a}(f) when $\alpha =0.0125$. 
This causes $P$ to grow transiently faster than in the nonlinear-skin regime, between the first slow down and the subsequent decay marking the boundaries of this regime, see blue curve in Fig.~\ref{fig:participation_ratio_nonrecip_weak_all}. 
Crucially, because these behaviors are absent in both the linear and nonlinear-skin regimes, they signal the onset of nonlinear eigenmode interactions, consistent with the wave-mixing regime illustrated in Fig.~\ref{fig:schematic}(c).

For times higher than this minimum of $P(t)$ marked by dotted horizontal lines in Fig.~\ref{fig:participation_ratio_nonrecip_weak_all}, the dynamics enters the third stage.
Within the latter, the $P$ exhibits a non-smooth time dependence and tends to increase at rates significantly larger than that observed in the nonlinear-skin regime.
Turning to the spatio-temporal evolution of the norm density profiles in Figs.~\ref{fig:heatmaps_real_space_01a}(b)–(d), we find that both the large amplitude  main pulse and the secondary waves lose their coherence and break into smaller localized structures, which appear to move chaotically within the excited region of the HN lattice.
The stochastic nature of the norm density evolution is also reflected in its mean position which shows random fluctuations in its time dependence as it shifts toward the direction favored by the NHSE, red curves in Figs.~\ref{fig:heatmaps_real_space_01a}(c)–(d).
Note that by numerically fitting $P(t)$~\cite{SMS2018} within the appropriate time windows, we find that $P \propto  t$ for all the representative nonlinear cases above as depicted in Fig.~\ref{fig:participation_ratio_nonrecip_weak_all}.
These results are consistent with the self-trapping regime (Sec.~\ref{subsec:selftrapping}), which is especially clear for $\alpha=100$ [magenta curve in Fig.~\ref{fig:participation_ratio_nonrecip_weak_all} and heatmap in Fig.~\ref{fig:heatmaps_real_space_01a}(d)].

We also verified numerically that the wavefront position and the second moment of the norm density distribution scale as $\sim t$ and $\sim t^2$, respectively.
These results confirm the ballistic nature of bulk wave-packet spreading within the self-trapping regime.
The corresponding plots are omitted here to avoid repetition.

With this in mind, the change in the concavity in the time dependence of $P$ likely signals the onset of nonlinear interactions, and its time of occurrence can be naturally associated with $T_{\mathrm{mix}}$.
It follows that $T_{\mathrm{mix}}$ is determined from the time at which $\ddot{P}\rvert_{t=T_{\mathrm{mix}}}$ changes sign.
In contrast, $T_{\mathrm{trap}}$ is identified from the lowest minimum of the participation number, $\dot{P}\rvert_{t=T_{\mathrm{trap}}}=0$.
One should keep in mind, however, that this extremum may be either local or global as shown in  Fig.~\ref{fig:participation_ratio_nonrecip_weak_all}.
From the data in Fig.~\ref{fig:participation_ratio_nonrecip_weak_all}, we find
$T_{\mathrm{mix}}=70$, $34$, $0$, and $0$ for $\alpha=0.0125$, $0.1$, $0.8$, and $100$, respectively.
Similarly, we obtain $T_{\mathrm{trap}}=126$, $94$, $50$, and $3$ for $\alpha=0.0125$, $0.1$, $0.8$, and $100$, respectively.
These times are indicated by the green dashed and black dotted horizontal lines in Figs.~\ref{fig:heatmaps_real_space_01a}(b)–(d), for $T_{\mathrm{mix}}$ and $T_{\mathrm{trap}}$ respectively.
Consequently, we observe that both $T_{\mathrm{mix}}$ and $T_{\mathrm{trap}}$ decrease as $\alpha$ increases, in agreement with the theoretical predictions of Sec.~\ref{subsec:time_scales}.

To further illustrate the different dynamical regimes of the wave-packets, we project the spatio-temporal evolution of the complex amplitude onto the eigenmodes, namely, $y_n(t) = \sum_{k} c_k(t) u_{k,n}$.
In this context, we focus on the representative cases shown in Fig.~\ref{fig:heatmaps_real_space_01a}: (a) $\alpha=0$, (b) $\alpha=0.0125$, (c) $\alpha=0.8$, and (d) $\alpha=100$.
For the linear case shown in Fig.~\ref{fig:heatmaps_pbc_space_01a}(a), the initial wave-packet at $t=0$ has a Gaussian profile centered at $k_0=0$.
As time evolves, the norm density profile is amplified while propagating in eigenmode space, with its center of mass $K(t)$ (Sec.~\ref{subsec:secular_form}) shifting toward larger wavenumbers.
This behavior is clearly illustrated in Fig.~\ref{fig:heatmaps_pbc_space_01a}(e), which displays snapshots of the amplitude evolution of Fig.~\ref{fig:heatmaps_pbc_space_01a}(a) at $t=20$ (blue), $t=95$ (red), $t=112$ (cyan), and $t=150$ (magenta).

Remarkably, as soon as nonlinearity is introduced, the computed values of $T_{\mathrm{mix}}$ and $T_{\mathrm{trap}}$ clearly separate the different dynamical regimes of the wave-packet, Figs.~\ref{fig:heatmaps_pbc_space_01a}(b)-(d).
Note that these characteristic times are indicated by green dashed and black dotted horizontal lines, corresponding to $T_{\mathrm{mix}}$ and $T_{\mathrm{trap}}$, respectively.
Within the nonlinear-skin regime, the dynamics remains predominantly coherent in eigenmode space, closely resembling the linear behavior observed for $\alpha=0$, as illustrated in Fig.~\ref{fig:heatmaps_pbc_space_01a}(b) for $\alpha=0.0125$.
By contrast, in the wave-mixing regime, the excited eigenmodes start to resonate with neighboring ones, leading to the emergence of secondary wave-packets [see Fig.~\ref{fig:heatmaps_pbc_space_01a}(b) for $\alpha = 0.0125$] or the widening of the main one clearly visible in Fig.~\ref{fig:heatmaps_pbc_space_01a}(c) for $\alpha=0.8$.

Beyond the wave-mixing regime, nonlinear resonances become increasingly strong, eventually exciting the entire eigenmode spectrum. 
These rapid resonant interactions manifest as stochastic fluctuations of the mean position of the norm density, as depicted by the red curves in Figs.~\ref{fig:heatmaps_pbc_space_01a}(b)–(d). 
Moreover, the excited eigenmodes attain very large amplitudes regardless of whether they are attenuated or amplified in the linear limit, Fig.~\ref{fig:heatmaps_pbc_space_01a}(f).
For all representative nonlinear cases, $T_{\mathrm{mix}}<T_{\mathrm{jump}}$, with $T_{\mathrm{jump}}$ lying beyond the final simulation time.
These numerical results therefore support the absence of a non-Hermitian wave jump in the presence of nonlinearity and are fully consistent with the theoretical predictions of Sec.~\ref{subsec:jump}.

\begin{figure}[!tb]
    \centering
    \includegraphics[width=\columnwidth]{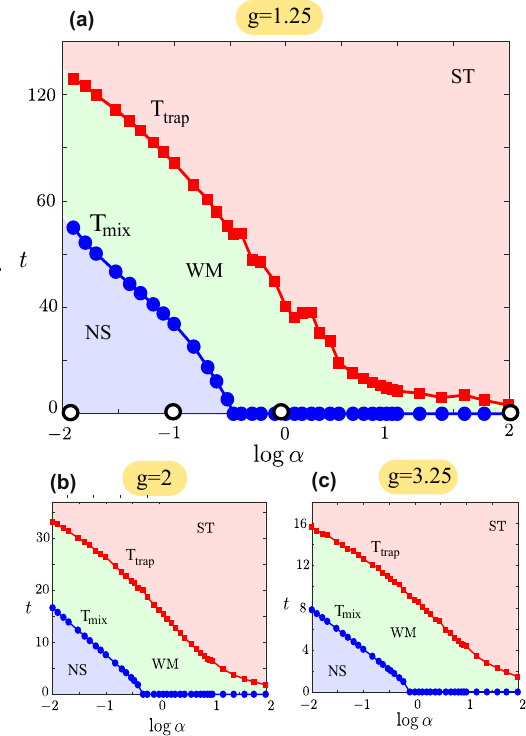}
    \caption{{\it Numerical results.} 
                Dynamical regimes of nonlinear wave-packets in the HN lattice.
                The diagram is shown in the $(\alpha,t)$ plane for different values of the nonreciprocity parameter:
                (a) weak nonreciprocity, $g=1.25$, (b) intermediate nonreciprocity, $g=2$, and (c) strong nonreciprocity, $g=3.25$.
                The colored regions indicate the three dynamical regimes identified in this work: nonlinear-skin (blue), wave-mixing (green), and self-trapping (red).
                Blue connected circles show the time $T_{\mathrm{mix}}$ at which $P(t)$ first deviates from its linear-like monotonic growth, whereas red connected squares indicate the time $T_{\mathrm{trap}}$ at which $P(t)$ reaches its first local minimum after $t=0$, see text for details.
                The white dots in (a) correspond, from left to right, to the representative cases $\alpha=0.0125$, $0.1$, $0.8$, and $100$.
                The dynamics of these cases are presented in Figs.~\ref{fig:participation_ratio_nonrecip_weak_all}, \ref{fig:heatmaps_real_space_01a}, and \ref{fig:heatmaps_pbc_space_01a}.
    }
    \label{fig:numerical_regime_diagram}
\end{figure}

\subsection{\label{subsec:num_diagram}Dynamical regime diagrams}
We repeat the calculations of $T_{\mathrm{mix}}$ and $T_{\mathrm{trap}}$ over the range $\alpha\in[10^{-2},10^{2}]$.
The results of these computations are presented in Figs.~\ref{fig:numerical_regime_diagram}(a)--(c) for the same HN chain considered above, with $g=1.25$, $g=2$, and $g=3.25$, respectively.
For each fixed value of the asymmetric coupling, both $T_{\mathrm{mix}}$ and $T_{\mathrm{trap}}$ decrease with increasing nonlinearity, until they saturate at zero or near-zero values.
Note that the parameter values corresponding to the representative cases shown in Figs.~\ref{fig:participation_ratio_nonrecip_weak_all}, \ref{fig:heatmaps_real_space_01a}, and \ref{fig:heatmaps_pbc_space_01a} are marked by white dots in Fig.~\ref{fig:numerical_regime_diagram}(a), ordered from left to right for $\alpha= 0.0125$, $\alpha= 0.1$, $\alpha=0.8$ and $\alpha=100$.
These representative cases were chosen to clearly illustrate the possible outcomes of the wave-packet.
In particular, when the nonlinear coefficient is sufficiently small like $\alpha=0.0125$ and $\alpha=0.1$ [the first two dots from the left in Fig.~\ref{fig:numerical_regime_diagram}(a)], the wave-packet is launched in the nonlinear-skin regime, leaving a sufficiently large time window for coherent wave dynamics to be observed, as in Figs.~\ref{fig:heatmaps_real_space_01a} and \ref{fig:heatmaps_pbc_space_01a}.
By contrast, for larger nonlinearities, the wave-packet may be launched directly into the wave-mixing or self-trapping regime, where incoherent dynamics dominates; see Figs.~\ref{fig:heatmaps_pbc_space_01a}(c) and \ref{fig:heatmaps_pbc_space_01a}(d).

Conversely, for fixed $\alpha$, both $T_{\mathrm{mix}}$ and $T_{\mathrm{trap}}$ decrease at comparable rates as nonreciprocity increases, as shown in Figs.~\ref{fig:numerical_regime_diagram}(a)--(c). 
Consequently, the wave-packet dynamics still passes through all three regimes regardless of the nonreciprocity strength. 
However, increasing nonreciprocity shortens the temporal windows associated with the nonlinear-skin and wave-mixing regimes, causing the system to enter the self-trapping regime at progressively earlier times. 
This behavior is consistent with the theory developed in Sec.~\ref{subsec:time_scales}.
Indeed, stronger nonreciprocity leads to faster exponential wave amplification, thereby enhancing nonlinear effects and driving the dynamics more rapidly toward its asymptotic self-trapped state, Fig.~\ref{fig:numerical_regime_diagram}(c).
In general, when the nonlinear coefficient of the medium cannot be tuned, the time interval over which nonlinear wave acceleration and other coherent pattern phenomena can be observed becomes increasingly narrow as the coupling asymmetry is strengthened.
It is worth emphasizing that we have verified the robustness of these numerical results above against changes in computational parameters, such as the integration time step.

\subsection{Nonlinear wave acceleration in the nonlinear-skin regime}

\begin{figure}[!tb]
    \centering
    \includegraphics[width=\columnwidth]{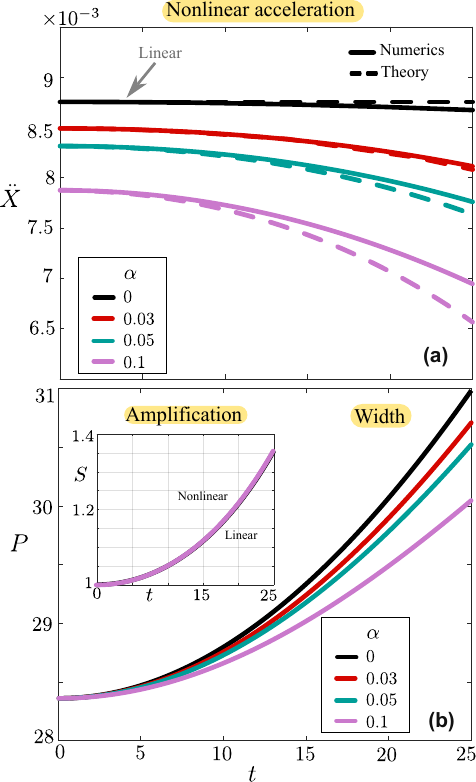}
    \caption{{\it Numerical results.}
            (a) Time evolution of the wave-packet acceleration $\ddot{X}$ within the nonlinear-skin (NS) regime shown in Fig.~\ref{fig:numerical_regime_diagram}(a). 
            The black curve corresponds to the linear case, while the colored curves represent the nonlinear cases with $\alpha = 0.03$ (red), $\alpha = 0.05$ (cyan), and $\alpha = 0.1$ (magenta). 
            The dashed lines show the corresponding analytical predictions for the acceleration dynamics. 
            For clarity, they were rescaled to match the values obtained from the lattice simulations at $t=0$.
            (b) Same as in (a), but for the dependence of the participation number $P(t)$ against time. 
            The inset shows the time evolution of the norm $S(t)$ (see text for details), for cases with $\alpha =0$ and $\alpha = 0.1$, colored according to the codes in panels (a) and (b).
    }
    \label{fig:numerical_acceleration_width}
\end{figure}

Having established the dynamical-regime diagrams for the wave-packet, we now investigate the effect of nonlinearity on its acceleration. We focus on the nonlinear-skin regime and consider a weakly nonreciprocal HN lattice with $g=1.25$, as shown in Fig.~\ref{fig:numerical_regime_diagram}(a).
Figure~\ref{fig:numerical_acceleration_width}(a) presents the numerically computed acceleration up to $t=25$ for $\alpha=0$, $0.03$, $0.05$, and $0.1$, represented by the black, red, cyan, and magenta solid curves, respectively. The corresponding analytical predictions for the time-dependent acceleration, obtained from Eq.~\eqref{eq:ode_nonlinear_width_acceleration_02}, are superimposed for comparison. To facilitate a direct comparison of their temporal evolution, the analytical curves are rescaled so that their initial values coincide with the corresponding numerical results at $t=0$. The differences in magnitude between the unscaled analytical and numerical accelerations are shown in Fig.~\ref{figapp:comparison_acceleration_theory_numerics} of Appendix~\ref{secapp:nonlinear_waves}.

In the linear case, the acceleration computed from the lattice dynamics initially takes the value $\ddot{X}=8.8\times10^{-3}$ and remains nearly constant throughout the evolution, reaching $\ddot{X}=8.67\times10^{-3}$ at the end of the simulation, as shown by the black solid curve. This behavior is in very good agreement with the theoretical prediction represented by the black dashed curve.

When weak focusing nonlinearities are introduced, the initial acceleration decreases as the nonlinear coefficient increases, in accordance with the analytical prediction. For the representative cases shown in Fig.~\ref{fig:numerical_acceleration_width}(a), the initial acceleration is $\ddot{X}=8.5\times10^{-3}$, $8.3\times10^{-3}$, and $7.87\times10^{-3}$ for $\alpha=0.03$, $0.05$, and $0.1$, respectively.
The subsequent evolution also displays the expected effect of focusing nonlinearity.
That is to say, both the analytical and numerical accelerations decrease with time, and the decay becomes more pronounced as $\alpha$ increases. 
At early times, the analytical and numerical curves exhibit nearly identical decay rates for all considered nonlinearities. 
At later times, however, the analytical theory tends to overestimate the rate of decay, with the discrepancy becoming more noticeable as $\alpha$ increases. 
Nevertheless, Fig.~\ref{fig:numerical_acceleration_width}(a) demonstrates good agreement between the analytical predictions and the numerical simulations.

Turning to the effect of nonlinearity on the wave-packet width, Fig.~\ref{fig:numerical_acceleration_width}(b) shows the time evolution of the participation number for the same simulations considered in Fig.~\ref{fig:numerical_acceleration_width}(a). As the nonlinear coefficient increases, the participation number grows more slowly, indicating a progressive suppression of wave-packet broadening. 
This behavior is consistent with the expected effect of focusing nonlinearities.
We further verify numerically that, within the nonlinear-skin regime, the total norm $S(t)$ exhibits nearly identical time dependence in the linear and nonlinear cases.
Indeed, the curves obtained for the representative cases $\alpha=0$ (black) and $\alpha=0.1$ (magenta) nearly overlap, as shown in the inset of Fig.~\ref{fig:numerical_acceleration_width}(b).

We performed the same analysis for defocusing nonlinearities. In this case, the analytical predictions and numerical simulations remain in good agreement, both showing an increase in the acceleration and a faster broadening of the initial wave-packet with time, consistent with the results of Sec.~\ref{subsec:nonlinear_acceleration}. We do not include these results in the manuscript, however, because they yield plots that are qualitatively similar to those obtained for focusing nonlinearities.

\section{\label{sec:conclusion}Conclusions and future challenges}

In conclusion, we have shown that even weak nonlinearity qualitatively reshapes wave-packet dynamics in the Hatano--Nelson (HN) lattice. 
Our theoretical analysis, supported by numerical simulations, reveals that a finite-amplitude wave-packet evolves successively through three dynamical regimes: nonlinear-skin, wave-mixing, and self-trapping. 
These regimes can be understood, and their boundaries estimated, by comparing the nonlinear frequency shift $\nu(t)$ with two characteristic spectral scales: the average eigenfrequency spacing $d$ and the spectral width $\Delta$. 
Importantly, we find that nonreciprocal amplification causes $\nu(t)$ to grow exponentially in time. 
As long as $\nu(t)<d$, nonlinearity remains perturbative, and the wave-packet propagates coherently within the nonlinear-skin regime.
When $d<\nu(t)<\Delta$, the wave-packet enters the wave-mixing regime, in which resonant interactions between eigenmodes lead to energy exchange and a loss of wave-packet coherence.
Finally, when $\nu(t)>\Delta$, self-trapping occurs: a fraction of the wave-packet becomes localized, while the remainder spreads ballistically in the direction favored by the asymmetric couplings.

Within the nonlinear-skin regime, our perturbative analysis shows that nonlinearity modifies both the magnitude and the time dependence of the wave-packet acceleration. 
Focusing nonlinearities suppress the acceleration and cause it to decrease with time, whereas defocusing nonlinearities enhance it and cause it to increase. 
These analytical predictions are corroborated by direct numerical simulations of the lattice model. 
We also showed that nonlinear interactions generally break up coherent wave-packet dynamics before the non-Hermitian wave jump can occur.

Our work highlights the crucial role of nonlinearity in nonreciprocal media and is relevant to a broad range of experimental platforms, including mechanical, optical, acoustic, atomic, and electrical systems, in which nonlinear effects naturally arise and may be exploited for wave control. More broadly, these results open several promising directions for future research. 
For instance, although we have established the occurrence of energy transfer between eigenmodes, a systematic characterization of the wave-mixing processes underlying nonreciprocal wave dynamics remains an open problem. 
Finally, extending the present framework to include dissipation~\cite{VGBVTCC2025,VGGSMC2024,JMAFS2025}, external driving, higher spatial dimensions, and more general forms of nonlinearity also represents an important direction for future work.

\begin{acknowledgments}
B.M.M. acknowledges partial support from the Israel Science Foundation (ISF) and the Bourses d’Accueil pour Chercheur Invit\'e (BACI) of the Laboratoire d’Acoustique de l’Universit\'e du Mans (LAUM) and the Institut d’Acoustique -- Graduate School (IA--GS).
V.A. acknowledges support from the EU H2020 research and innovation programme under ERC Starting Grant “NASA” (Grant Agreement No. 101077954).
We also thank the two anonymous referees for their constructive comments and suggestions, which helped improve the clarity and presentation of this work.
\end{acknowledgments}

\appendix

\section{\label{secapp:continuum}Continuum limit and partial differential equation (PDE) of motion}
Starting from the system of ordinary differential equations (ODEs) describing all oscillators, $y_n$ in the HN lattice, we apply the continuum limit approximation, setting $x=n$, $y_n = y(x)$ and 
\begin{equation}
    y_{n\pm1} = y \pm \frac{\partial y}{\partial x} + \frac{1}{2}\frac{\partial^2y}{\partial x^2} + \ldots 
\end{equation}
It follows that
\begin{equation}
    -i \frac{\partial y}{\partial t} = - \varepsilon_0  y - b \frac{\partial y}{\partial x} + D \frac{\partial^2 y}{\partial x^2} + \alpha \lvert y \rvert^2 y
\end{equation}
when considering terms up to the second derivative~\cite{R1999}.
Here, the $\varepsilon_0 = -(g+1)$, $b= g-1$  and $D = (g+1)/2$.
A simple transformation $y \rightarrow y e^{i \varepsilon_0 t}$
leads to
\begin{equation}
    i \frac{\partial y}{\partial t} =  b \frac{\partial y}{\partial x} - D \frac{\partial^2 y}{\partial x^2} - \alpha \lvert y \rvert^2 y.
\end{equation}
This is the PDE of motion shown in the main text.

\section{\label{secapp:nonlinear_waves}Analytical derivation of the nonlinear acceleration and width}
In this Appendix, we outline the main steps leading from Eqs.~\eqref{eq:var_param_1} to~\eqref{eq:ode_nonlinear_width_acceleration_02}.
These equations govern the evolution of the parameters, $A$, $\sigma$, $\Phi$, $\rho$ and $\eta$ of the trial solution 
\begin{equation}
    y(\xi, t) = A e^{-\frac{\xi^2}{4\sigma^2}+ i\Phi \xi^0 i\rho\xi^1 + i\eta\xi^2}, \quad  \xi = x - X,
    \label{eqapp:ansatz}
\end{equation}
introduced in Eq.~\eqref{eq:ansatz_nonlinear} of the main text.
It follows that 
\begin{equation}
    s = \lvert y \rvert^2 = A^2 e^{-\frac{\xi^2}{2\sigma^2}},
    \label{eqapp:presub_01}
\end{equation}
when calculating its norm density (distribution).
Further, using Eq.~\eqref{eqapp:ansatz}, we compute the space derivatives of the complex amplitude, $y$,
\begin{equation}
    \frac{1}{y}\frac{\partial y}{\partial x} = i \rho + 2i\eta \xi - \frac{\xi}{2\sigma^2}
    \label{eqapp:presub_02}
\end{equation}
and,
\begin{equation}
    \begin{split}
        \frac{1}{y}\frac{\partial^2 y}{\partial x^2} &= - \rho^2 - 4\rho \eta \xi - \frac{i\rho \xi}{\sigma^2} - \frac{2i\eta \xi^2}{\sigma^2} \\
        & - 4 \eta^2 \xi^2 + \frac{\xi^2}{4\sigma^4} + 2i\eta - \frac{1}{2\sigma^2}.
    \end{split}
    \label{eqapp:presub_03}
\end{equation}
Turning now to its time derivative, we have 
\begin{equation}
    \begin{split}
        \frac{1}{y} \frac{\partial y}{\partial t} &= \frac{\dot{A}}{A} + i\Phi + i\dot{\rho} \xi - i\rho \dot{\xi} + i\dot{\eta} \xi^2 \\
        & - 2i\eta \dot{X} \xi + \frac{\dot{X}\xi}{2\sigma^2} + \frac{\dot{\sigma}\xi^2}{2\sigma^3},
    \end{split}
    \label{eqapp:presub_04}
\end{equation}
where we use the fact that $\dot{\xi} = - \dot{X}$.

Substituting Eqs.~\eqref{eqapp:presub_01} to~\eqref{eqapp:presub_04} into the PDE of motion,
\begin{equation}
    \frac{i}{y}\frac{\partial y}{\partial t} =   \frac{b}{y}\frac{\partial y}{\partial x} - \frac{D}{y}\frac{\partial^2 y}{\partial x^2} - \alpha \lvert y \rvert^2,
    \label{eqapp:pde_of_motion_divided_y}
\end{equation}
we find the following for the imaginary part of the collective coordinate equation,
\begin{equation}
    \begin{split}
        \left(\frac{\dot{A}}{A} + 2 D \eta - b\rho \right) & + \left(\frac{\dot{X}}{2\sigma^2}  - 2b\eta + \frac{D\rho}{\sigma^2} \right) \xi \\
        & + \left(\frac{\dot{\sigma}}{2\sigma^3} - \frac{2D\eta}{\sigma^2} \right) \xi^2 = 0 .
    \end{split}
    \label{eqapp:imag}
\end{equation}
Multiplying Eq.~\eqref{eqapp:imag} by $s$ and integrating over $x$, we obtain
\begin{equation}
    S \left(\frac{\dot{A}}{A} + 2 D \eta - b\rho \right) + S\sigma^2 \left(\frac{\dot{\sigma}}{2\sigma^3} - \frac{2D\eta}{\sigma^2} \right)  = 0.
\end{equation}
Note that along the way, we made great use of the expressions of the moments of Gaussian wave-packets of Appendix~\ref{secapp:moments}.
It follows that 
\begin{equation}
    \frac{\dot{A}}{A} + 2 D \eta - b\rho = 0,
    \label{eqapp:dot_param_01}
\end{equation}
and 
\begin{equation}
    \dot{\sigma} = 4 D \eta \sigma.
    \label{eqapp:dot_param_02}
\end{equation}
Clearly, since $S = \sqrt{2\pi} A^2 \sigma$, we have 
\begin{equation}
    \frac{\dot{S}}{S} = 2 \frac{\dot{A}}{A} + \frac{\dot{\sigma}}{\sigma},
\end{equation}
leading to 
\begin{equation}
    \dot{S} = 2b\rho S,
\end{equation}
which is also used in the main text.
Further, multiplying Eq.~\eqref{eqapp:imag} by $\xi$ and integrating over $x$, we find (Appendix~\ref{secapp:moments})
\begin{equation}
    S\sigma^2 \left(\frac{\dot{X}}{2\sigma^2}  - 2b\eta + \frac{D\rho}{\sigma^2} \right) = 0.
\end{equation}
It follows that,
\begin{equation}
    \dot{X} = 2D\rho + 4 b\eta \sigma^2.
    \label{eqapp:dot_param_03}
\end{equation}

On the other hand, the real part of the collective coordinate equation leads to,
\begin{equation}
    \begin{split}
        & \left(2\eta \dot{X} - \dot{\rho} + \frac{b}{2\sigma^2} - 4D\rho \eta \right) \xi + \left(-\dot{\eta} - 4D\eta^2 + \frac{D}{4\sigma^4} \right) \xi^2 \\
         &-\dot{\Phi} - \rho \dot{X} - D\rho^2 - \frac{D}{2\sigma^2} + \alpha s = 0 .
    \end{split}
    \label{eqapp:real_part_eq}
\end{equation}
We multiply Eq.~\eqref{eqapp:real_part_eq} by $\xi s$ and integrate over $x$.
This results in
\begin{equation}
    S\sigma^2 \left( 2\eta \dot{X} - \dot{\rho} + \frac{b}{2\sigma^2} - 4D\rho \eta \right) = 0.
\end{equation}
Consequently, we find
\begin{equation}
    \dot{\rho} = \frac{b}{2\sigma^2} + 8 b\eta^2 \sigma^2,
    \label{eqapp:dot_param_04}
\end{equation}
when also considering Eq.~\eqref{eqapp:dot_param_03}.
Once again, one need to make use of the expressions of the moments in Appendix~\ref{secapp:moments}.

\begin{figure}[!tb]
    \centering
    \includegraphics[width=\columnwidth]{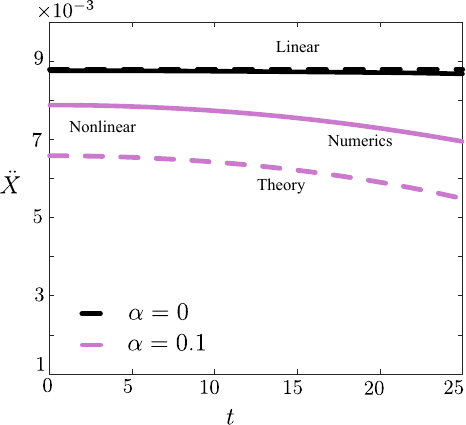}
    \caption{
            Comparison of the magnitudes of the wave-packet acceleration, $\ddot{X}$, obtained from the theory (dashed curves) and numerical lattice simulations (solid curves).
            The black curve corresponds to the linear case, $\alpha=0$, whereas the magenta curves denote a nonlinear counterpart with $\alpha=0.1$.
            The lattice consists of $N=512$ sites with OBCs and $g=1.25$, and the parameters of the initial Gaussian pulse are fixed to $\sigma_0=8$, $k_0=0$, and $n_0=0$, see also Fig.~\ref{fig:numerical_acceleration_width}.
    }
    \label{figapp:comparison_acceleration_theory_numerics}
\end{figure}

To eliminate the constant terms, which mainly contribute to the phase equation, $\dot{\Phi}$, we multiply Eq.~\eqref{eqapp:real_part_eq} by $(\xi^2 - \sigma^2)s$ and integrating over $x$.
We find that (Appendix~\ref{secapp:moments}),
\begin{equation}
      2S \sigma^4 \dot{\eta} -\frac{DS}{2} + 8 DS \eta^2 \sigma^4 + \frac{\alpha S^2 \sigma}{4\sqrt{\pi}} = 0.
\end{equation}
Dividing this equation by $-2S\sigma^4$ we obtain
\begin{equation}
    \dot{\eta} = \frac{D}{4\sigma^4} - 4D\eta^2 - \frac{\alpha S}{8\sqrt{\pi}}{\sigma^3}.
\end{equation}
This is the expression used in the main text.
\paragraph{Nonlinear width.}
To deduce the second order equation of the width, we need to differentiate $\dot{\sigma}=4D\eta \sigma$.
It follows that,
\begin{eqnarray}
    \ddot{\sigma} &=& 4 D \left(\dot{\eta}\sigma + \eta \dot{\sigma} \right) \\
    &=& \frac{D^2}{\sigma^3} - \frac{\alpha D S}{2\sqrt{\pi}\sigma^2}.
\end{eqnarray}

\paragraph{Nonlinear acceleration.}
Similarly for the center of mass motion, we differentiate $\dot{X} = 2D\rho + 4 b\eta \sigma^2$, following
\begin{eqnarray}
    \ddot{X} &=& 2D\dot{\rho} + 4b \left(\dot{\eta}\sigma^2 + 2\eta \sigma \dot{\sigma} \right),\\
    &=& \frac{2bD}{\sigma^2} + \frac{2b}{D} \dot{\sigma}^2 - \frac{\alpha b S}{2\sqrt{\pi}\sigma}.
\end{eqnarray}
where we use the fact that $32bD\eta^2 \sigma^2 = 2b\dot{\sigma}^2/D$.
It follows that we obtain the acceleration of the center of mass of the Gaussian wave-packet.

For completeness, we present the method to find the phase equation, $\dot{\Phi}$.
Substituting, 
\begin{equation}
    \alpha s = \alpha A^2 e^{-\frac{\xi^2}{2\sigma^2}} = \alpha A^2 \left( 1 -\frac{\xi^2}{2\sigma^2} \right).
    \label{eqapp:expansion_nonlinear}
\end{equation}
 into Eq.~\eqref{eqapp:real_part_eq} and multiplying the latter by $\xi^2 s$ then integrating it over $x$, it follows that
\begin{equation}
    \dot{\Phi} = \rho \dot{X} - D\rho^2 - \frac{D}{2\sigma^2} +  \frac{\alpha S}{\sqrt{2\pi}}.
\end{equation}

Figure~\ref{fig:heatmaps_pbc_space_01a} compares the analytical predictions (dashed curves) with direct lattice simulations (solid curves) for the acceleration of an initial Gaussian wave-packet with $\sigma_0=8$, $k_0=0$, and $n_0=0$, in a lattice of $N=512$ sites with OBCs and $g=1.25$. 
We show both the linear case, $\alpha=0$ (black), and a nonlinear case, $\alpha=0.1$ (magenta), up to $t=25$, which lies within the nonlinear-skin regime of the nonreciprocity strength above, Fig.~\ref{fig:numerical_regime_diagram}(a). 
Overall, both the theory and the lattice simulations are in good agreement.
In the linear limit, the wave-packet acceleration remains nearly constant in time, validating the weakly dispersive PDE approximation. 
In the presence of nonlinearity, the acceleration decreases with time, with the analytical values remaining below those obtained from the lattice simulations. 
Nevertheless, this offset is small relative to the acceleration magnitude even at such a large nonlinear strength ($\alpha=0.1$), and both approaches predict nearly identical decay rates.
Moreover, as $\alpha\rightarrow 0$, the offsets of the analytical and numerical results tends to vanish.

\section{\label{secapp:moments}Some useful moments of the Gaussian wave-packet}
In this section, we provide explicit formulas for several moments of the distribution and of the position for a Gaussian wave-packet. These moments are used in Appendix~\ref{secapp:nonlinear_waves}.
Considering the distribution
\begin{equation}
    s = \lvert y \rvert^2 = A^2 \exp \left( -\frac{\xi^2}{2\sigma^2} \right),
\end{equation}
we obtain 
\begin{equation}
    \int s dx = S, \mbox{ and } \int \xi^\ell s dx = 0, \mbox{ with } \ell = 1, 3, 5, \ldots .
\end{equation}
The latter is true since the integrands are odd functions.
Further,
\begin{equation}
    \int \xi^2 s dx = S\sigma^2, \mbox{ and } \int \xi^4 s dx = 3S \sigma^4.
\end{equation}

On the other hand, assuming the squared distribution,
\begin{equation}
    s^2 = A^4 \exp \left( -\frac{\xi^2}{\sigma^2} \right),
\end{equation}
we find that
\begin{equation}
    \int s^2 dx = \frac{S^2}{2\sqrt{\pi}\sigma}, \mbox{ and }\int \xi^2 s^2 dx = \frac{S^2\sigma}{4\sqrt{\pi}}.
\end{equation}
These expressions can be found using standard Gaussian integral formulas~\cite{AS1964}.



\bibliography{apssamp}

\begin{thebibliography}{78}
\bibitem [{KIK()}]{KIKIPO}%
  \BibitemOpen
  \href@noop {} {}\bibinfo {note} {The eigenmodes are orthonormalized.}\BibitemShut {Stop}%
\end{thebibliography}

\end{document}